\documentclass[preprint,aps,12pt,showpacs,nofootinbib,tightenlines]{revtex4}
\usepackage{mathrsfs}
\usepackage{amsmath}
\usepackage{amssymb}
\usepackage{epsfig}
\usepackage{epstopdf}
\usepackage{graphicx}
\usepackage{subfigure}
\usepackage{booktabs}
\usepackage{float}
\usepackage{amsmath}
\usepackage{color}
\usepackage{multirow}

\def\be{\begin{eqnarray}}
\def\en{\end{eqnarray}}
\def\non{\nonumber\\}

\begin{document}
%%--------------------------------------------
\title{Semi-leptonic $B$ decays to tensor mesons}
\author{Shao-Qin Guo, Zhi-Qing Zhang\footnote{ zhangzhiqing@haut.edu.cn (corresponding author)}, Xin-Yu Cai and Feng-Qing Hu
  } %%
\affiliation{ \it \small   School of Physics and Advanced Energy, Henan University of Technology, Zhengzhou, Henan 450001, China}
\date{\today}
\begin{abstract}
Using the form factors of the transtions $B\to T$ with $T$ refering to a tensor meson, such as $a_2(1320), f_2(1270),K^*_2(1430), D_2^*(2460)$ and $D^*_{2s}(2573)$, within the covariant light-front quark model (CLFQM), we
provide a detailed investigation of the corresponding semi-leptonic decays $B\to T\ell\nu_\ell$ with $\ell=e,\nu,\tau$. All the branching ratios of these
decays are larger than $10^{-5}$, in which the maximum value can reach up to $10^{-3}$, indicating promising prospects for experimental observation. Furthermore, we also calculate the longitudinal polarization fractions $f_L$ and forward-backward asymmetries $A_{FB}$ for these considered decays. All the decays $B\to T \ell\nu_{\ell}$ are dominated by the longitudinal
polarization, where the polarization fractions can reach up to $\sim70\%$ for the decays $B\to T \ell^{\prime}\nu_{\ell^\prime}$ with $\ell^\prime=e, \mu$, those of the decays $B\to T \tau\nu_{\tau}$ are a little smaller. The $A_{FB}$ values of the decays $B\to T \ell^{\prime}\nu_{\ell^\prime}$ and $B\to T \tau\nu_{\tau}$ have opposite signs.

%This paper employs the covariant light-front quark model to systematically calculate the transition form factors for $B$ mesons to light tensor mesons ($T$) and the observables for the semi-leptonic decays $B \to T \ell \nu_\ell$. Our form factor results show excellent agreement with other light-front models but exhibit differences compared to those from perturbative QCD (PQCD), reflecting the varying emphases of different methods in handling non-perturbative effects. The calculations indicate that the branching ratios for most decay channels are on the order of $10^{-4}$, making them promising for observation at Super B factories and LHCb. The decay products are predominantly longitudinally polarized, with a longitudinal polarization fraction of approximately $55\%\sim75\%$. The forward-backward asymmetry $A_{FB}$ is sensitive to lepton mass, being negative for electron or muon final states and positive for $\tau$ final states. These results provide important theoretical references for future experimental measurements.

\end{abstract}

 \vspace{1cm}

\maketitle

%=======================================================================
%                     Introduction
%=======================================================================
\section{Introduction}\label{intro}
The $B$ meson semi-leptonic decays play an important role in measuring the Cabibbo-Kobayashi-Maskawa (CKM)  matrix elements, testing the Standard Model (SM), understanding the dynamic mechanisms of weak decay, and investigating the new physics (NP) effects.
In this work, using the form factors of the transitions $B\to T$, we investigate the branching ratios, the longitudinal polarization fractions $f_L$ and forward-backward asymmetries $\mathcal{A}_{FB}$ for
the semi-leptonic decays $B\to T\ell\nu_\ell$. $B$ meson decay is a multi-scale problem, spanning from the electroweak scale ($m_W$) down to the hadronic scale ($\Lambda_{\text{QCD}}$), and the core difficulty lies in calculating transition matrix elements involving the strong interaction. The hadronization process inevitably involves non-perturbative Quantum Chromodynamics (QCD) mechanisms, introducing unavoidable theoretical uncertainties.

To address this non-perturbative challenge, theoretical physicists have developed a series of factorization schemes and phenomenological models. Among the various non-perturbative approaches, we will focus on the covariant light-front quark model (CLFQM) to deal with the $B\to T$ transition form factors. The CLFQM has some unique advantages compared
with other quark models. First, the light-front wave functions used in this approach are independent of the hadron momentum and thus are manifestly Lorentz invariant.
Moreover, the hadron spin is correctly constructed using the so-called Melosh rotation. Second, this model provides a relativistic treatment of the hadron. Since the final state meson
at the maximum recoil point ($q^2=0$) or in heavy-to-light transitions can be highly relativistic, considering the relativistic effects is crucial. Finally, the spurious contribution in the CLFQM, which depends
on the light-front orientation, is canceled by the zero-mode contribution, thereby restoring the covariance of the current matrix elements lost in the previous standard light-front quark
model. The CLFQM has been successfully extended to investigate $B$ decays
with S-wave mesons involved in the final states \cite{Sun:2025yiz,Yang:2025zvo,Yang:2024qij,Yang:2025gfz,2410,zhangzq}.

Based on the $SU(3)$ symmetry, the light P-wave tensor mesons with $J^P=2^+$ form a $^3P_2$ nonet, which involves the iso-vector mesons $a_2(1320)$, iso-doublet states $K^*_2(1430)$, and two iso-singlet mesons $f_2(1270)$ and $f^\prime_2(1525)$ \cite{wang-wei:1008,1997}. They have been well established in experiments. The $q\bar q$ quark component for the $a_2(1320)$and $K^*_2(1430)$ mesons is unambiguous. However, the quark composition of the iso-singlet states $f_2(1270)$ and $f^\prime_2(1525)$
is not entirely certain \cite{2008,2001}. It is generally considered that their wave functions exhibit mixing with a small mixing angle, and $f_2(1270)$ is primarily $(u\bar u+d\bar d)/\sqrt{2}$, while $f_2^\prime(1525)$ is dominantly made of $s\bar s$ \cite{cheng2010}. In the charmed meson spectrum, $D^{*}_2(2460)$ and $D^{*}_{s2}(2573)$ are tensor mesons with spin-parity $J^P = 2^+$, which are typically interpreted in the quark model as P-wave excited states with orbital angular momentum $L = 1$. In 1989, $D^{*0}_2(2460)$ was firstly observed by TPS Collaboration in the $D^+\pi^-$ invariant mass distribution \cite{334}, and its spin-parity $J^P = 2^+$ was determined by ARGUS collaboration through angular momentum analysis \cite{345}. Its charged partner, $D^{*}_2(2460)^+$, was soon discovered by ARGUS collaboration in the $D^{*}_2(2460)^+\to D^0\pi^+$ channel \cite{335} and confirmed by many other experimental collaborations including BaBar and Belle \cite{BaBar1,BaBar2,Belle}. Five years later, $D^*_{s2}(2573)$ was found by CLEO collaboration in the $D^0K^+$ invariant mass spectrum \cite{362} and confirmed by BaBar \cite{BaBar3} and LHCb \cite{LHCb}. It is worth noting that the width of $D^{*0}_2(2460)$ is about $2.8$ times as large as that of $D^*_{s2}(2573)$ \footnote{From now on, we will use $a_2, f_2, K^*_2, D^*_2$ and $D^{*}_{s2}$ to represent $a_2(1320),f_2(1270),K^*_2(1430), D^{*}_2(2460)$ and $D^{*}_{s2}(2573)$, respectively, for simplicity.}.

The outline of this paper is as follows. Section \ref{form} we introduce the formalism of the CLFQM, the hadronic matrix elements, and the helicity amplitudes via related form factors. Our results for the semi-leptonic decays $B \to T\ell\nu_\ell$, including the branching ratios, the longitudinal polarization fractions $f_L$, and the forward-backward asymmetries $\mathcal{A}_{FB}$, along with detailed discussions and comparisons with other theoretical predictions, are presented in Section \ref{numer}. Section IV comes to our conclusions. The complete analytical expressions, numerical results and their $q^2$ dependence for the $B \to T$ transition form factors are collected in Appendices A and B.

\section{Formalism}\label{form}
\subsection{Covariant Light-Front Quark Model}
\begin{figure}[H]
\centering \subfigure{
\begin{minipage}{5cm}
\centering
\includegraphics[width=5cm]{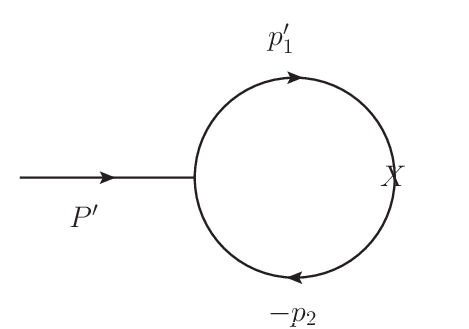}
\end{minipage}}
\subfigure{
\begin{minipage}{6cm}
\centering
\includegraphics[width=6cm]{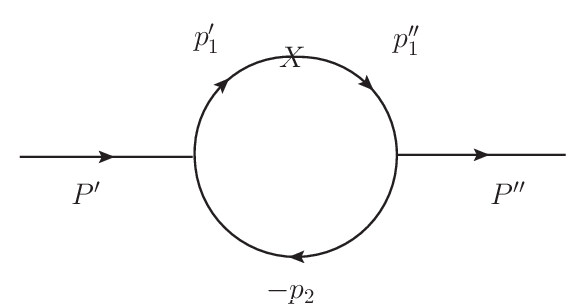}
\end{minipage}}
\caption{Feynman diagrams for $B$ decay (left) and transition
(right) amplitudes, where $P^{\prime(\prime\prime)}$ is the
incoming (outgoing) meson momentum, $p^{\prime(\prime\prime)}_1$
is the quark momentum, $p_2$ is the anti-quark momentum. }
\label{feyn}
\end{figure}

The Feynman diagrams for $B$ meson decay and transition amplitudes are shown in Figure \ref{feyn}.
In the CLFQM, momentum $p$ is conventionally parameterized via $p=(p^-,p^+,p_\perp)$ with
$p^\pm=p^0\pm p_z$, $p^2=p^+p^--p^2_\perp$.
The incoming (outgoing) meson has the mass $M^\prime(M^{\prime\prime})$
and the momentum $p^\prime=p_1^\prime+p_2 (p^{\prime\prime}=p_1^{\prime\prime}+p_2)$, where $p_{1}^{\prime(\prime\prime)} $
and $p_{2}$ are the momenta of the quark and anti-quark
inside the incoming (outgoing) meson with the mass $m_{1}^{\prime(\prime\prime)} $and $m_{2}$, respectively.
These momenta can be expressed in terms of the internal variables $(x_{i},p{'}_{\perp})$ as
\be
p_{1,2}^{\prime+}=x_{1,2} P^{\prime+}, \quad p_{1,2 \perp}^{\prime}=x_{1,2} P_{\perp}^{\prime} \pm p_{\perp}^{\prime},
\en
where $x_{1}+x_{2}=1$. Using these internal variables,
we can define some quantities for the incoming meson which will be used in the following calculations
\be
M_{0}^{\prime 2} &=&\left(e_{1}^{\prime}+e_{2}\right)^{2}=\frac{p_{\perp}^{\prime 2}+m_{1}^{\prime 2}}{x_{1}}
+\frac{p_{\perp}^{2}+m_{2}^{2}}{x_{2}}, \quad \widetilde{M}_{0}^{\prime}=\sqrt{M_{0}^{\prime 2}-\left(m_{1}^{\prime}-m_{2}\right)^{2}},\\
e_{i}^{(\prime)} &=&\sqrt{m_{i}^{(\prime) 2}+p_{\perp}^{\prime 2}+p_{z}^{\prime 2}}, \quad \quad p_{z}^{\prime}
=\frac{x_{2} M_{0}^{\prime}}{2}-\frac{m_{2}^{2}+p_{\perp}^{\prime 2}}{2 x_{2} M_{0}^{\prime}}. \label{m0p}
\en
It is similar to the case of the outgoing meson but with the superscript $\prime$ replaced by $\prime\prime$.

The Bauer-Stech-Wirble (BSW) form factors for the transition $B \to T$ are defined as follows
\be
\langle T(p^{\prime\prime} , \varepsilon) |\, \left. V_{\mu} \right. \,| B(p^\prime) \rangle &=& -i\frac{2}{m_{B}+m_T}\epsilon_{\mu\nu\alpha\beta}e^{*\nu}p^{\prime\alpha}p^{\prime\prime\beta}V^{BT}(q^2),
\\
\langle T(p^{\prime\prime} , \varepsilon) |\, \left. A_{\mu} \right. \,| B(p^{\prime}) \rangle &=& \left. 2 m_{T}\frac{e^{*}\cdot q}{q^2}q_\mu A^{BT}_{0}(q^2)+(m_{B}+m_T)\bigg[e^{*}_{\mu}
-\frac{e^{*}\cdot q}{q^2}q_\mu  \bigg] A^{BT}_{1}(q^2)
\right.\notag\\&&\left.
-\frac{e^{*}\cdot q}{m_B+m_V}\bigg[P_{\mu}
-\frac{m^2_{B}-m^2_{T}}{q^2}q_\mu  \bigg]A^{BT}_{2}(q^2)  \right.,
\en
where $P=p^\prime+p^{\prime\prime}$, $q=p^\prime-p^{\prime\prime}$ and $e^{*\mu}\equiv \varepsilon^{*\mu\nu}p^\prime_{\nu}/m_B$.  Throughout the paper we have adopted the convention $\epsilon_{0123} = -1$.
In the Isgur-Scora-Grinstein-Wise (ISGW) model, the general expression for the $B\to T$
transition is parametrized as
\be
\langle T(p^{\prime\prime} , \varepsilon) |\, \left. (V-A)_{\mu} \right. \,| B(p^\prime) \rangle &=& \left.
i h(q^2)\epsilon_{\mu\nu\rho\sigma}\varepsilon^{*\nu^\alpha} p^{\prime}_{\alpha}P^{\rho}q^{\sigma} -k(q^2)\varepsilon^*_{\mu\nu}p^{\prime\nu}
\right.\notag\\&&\left.
-b_+(q^2)\varepsilon^*_{\alpha\beta}p^{\prime\alpha}p^{\prime\beta}P
-b_-(q^2)\varepsilon^*_{\alpha\beta}p^{\prime\alpha}p^{\prime\beta}q  \right.,
\en
where the form factor $k(q^2)$ is dimensionless, and the dimensions of $h(q^2)$, $b_+(q^2)$ and $b_{-}(q^2)$ are $-2$. The polarization tensor, which satisfies $\epsilon_{\mu\nu} p^{\prime\prime\nu} = 0$, is symmetric and traceless. The spin-2 polarization tensors $\epsilon_{\mu\nu}$ can be constructed by use of the appropriate Clebsch-Gordan coefficients \cite{PDG} as follows
\be
\epsilon_{\mu\nu}(p,\pm2) &=& \epsilon_{\mu}(\pm)\epsilon_{\nu}(\pm), \quad \epsilon_{\mu\nu}(p,\pm1) = \frac{1}{\sqrt{2}}[\epsilon_{\mu}(\pm)\epsilon_{\nu}(0)+\epsilon_{\nu}(\pm)\epsilon_{\mu}(0)],\\
\epsilon_{\mu\nu}(p,0) &=& \frac{1}{\sqrt{6}}[\epsilon_{\mu}(+)\epsilon_{\nu}(-)+\epsilon_{\nu}(+)\epsilon_{\mu}(-)]+
\sqrt{\frac{2}{3}}\epsilon_{\mu}(0)\epsilon_{\nu}(0),
\en
where $\epsilon_\mu(0),\epsilon_\mu(\pm)$ are the polarization vectors for a massive vector state moving along the z-axis and can be given as
\be
\epsilon(0) &=& \frac{1}{m_{T}}(|\, \vec{p}_{T}\,|,0,0,E_{T}),\non
\epsilon(\pm) &=& \frac{1}{\sqrt{2}}(0,\mp 1,-i,0),
\en
with $E_T$ and $\vec{p}_T$ being the energy and momentum magnitude of the tensor meson in the $B$ meson rest frame, respectively, and $m_T$ denotes the mass of the tensor meson \cite{wang-wei:0901}.

The relations between these two different sets of form factors are
\be
V^{BT}(q^{2}) &=& m_{B}(m_{B}+m_{T})h(q^{2}),\\
A^{BT}_{1}(q^{2}) &=& \frac{m_{B}}{m_{B}+m_{T}}k(q^{2}),\\
A^{BT}_{2}(q^{2}) &=& -m_{B}(m_{B}+m_{T})b_{+}(q^{2}),\\
A_{0}^{BT}(q^{2}) &=& \frac{m_{B}}{2 m_{T}}[k(q^{2})+(m^{2}_{B}-m^{2}_{T})b_{+}(q^{2})+q^{2}b_{-}(q^{2})].
\en

The decay amplitude at the lowest order for the general $P\to T$ transition can be written as
\be
\mathcal{A}^{PT} &=& -i^{3} \frac{N_{c}}{(2\pi)^{4}} \int d^{4} p_{1}^{\prime} \frac{H_{P}^{\prime}(i H_{T}^{\prime\prime})}{N_{1}^{\prime}N_{1}^{\prime\prime}N_{2}} S_{\mu}^{ PT },
\en
where $N_1^{\prime(\prime\prime)} = p^{\prime(\prime\prime)2}_1 - m^{\prime(\prime\prime)2}_1$, $N_2 = p^2_2 - m^2_2$. The expression of $S^{PT}_\mu$ can be derived using the Lorentz contraction
\be
S^{PT}_\mu &=& \text{Tr}[(\not{p}^{\prime\prime}_1 + m^{\prime\prime}_1) \gamma^\mu \gamma_5(\not{p}^\prime_1 + m^\prime_1)\gamma_5(-\not{p}_2 + m_2)]
\non
&=& \left. 2 p^\prime_{1\mu}[M^{\prime2}+M^{\prime\prime2}-q^2-2 N_2-(m^\prime_1-m_2)^2-(m^{\prime\prime}_1+m_2)^2+(m^\prime_1+m^{\prime\prime}_1)^2]
\right.\notag\\&&\left.
+q_\mu[q^2-2 M^{\prime2}+N^\prime_1-N^{\prime\prime}_1+2 N_2+2(m^\prime_1-m_2)^2-(m^\prime_1+m^{\prime\prime}_1)^2]
\right.\notag\\&&\left.
+P_\mu[q^2-N^\prime_1-N^{\prime\prime}_1-(m^\prime_1+m^{\prime\prime}_1)^2\right.].
\en

In practice, we use the light-front decomposition of the Feynman loop momentum and integrate out the minus component through the contour method. If the covariant vertex functions are not singular when performing integration, the transition amplitudes will pick up the singularities in the anti-quark propagators. The integration then leads to
\be
N^{\prime(\prime\prime)}_1 \to \hat{N}^{\prime(\prime\prime)}_1 &=& x_1(M^{\prime(\prime\prime)2} - M^{\prime(\prime\prime)2}_0), \\
\int \frac{d^4 p^\prime_1}{N^\prime_1 N^{\prime\prime}_1 N_2} H^\prime_P H^{\prime\prime}_T S^{PT} &\to& -i\pi \int \frac{d x_2 d^2 p^\prime_\perp}{x_2 \hat{N}^\prime_1
\hat{N}^{\prime\prime}_1} h^\prime_P h^{\prime\prime}_T \hat{S}^{PT},\\
W^{\prime\prime}_T \to w^{\prime\prime}_T&=&M^{\prime\prime}_0+m^{\prime\prime}_1+m_2,
\en
with $p^{\prime\prime}_\perp = p^\prime_\perp - x_2 q_\perp$. The explicit forms of $h^\prime_P$ and $h^{\prime\prime}_T$ used in this work are given by
\be
h^\prime_P &=& (M^{\prime2} - M^{\prime2}_0) \sqrt{\frac{x_1 x_2}{N_c}} \frac{1}{\sqrt{2} \widetilde{M}^\prime_0} \varphi^\prime, \\
h^{\prime\prime}_T &=& (M^{\prime\prime2}- M^{\prime\prime2}_0) \sqrt{\frac{x_1 x_2}{N_c}} \frac{1}{\sqrt{2} \widetilde{M}^{\prime\prime}_0} \varphi^{\prime\prime}_p,
\en
where $\varphi^\prime$ and $\varphi^{\prime\prime}_p$ are the light-front wave function for s-wave (initial state) and p-wave (final state) mesons, respectively. Here, the phenomenological Gaussian-type wave functions are adopted in the following calculations,
\be
\varphi^{\prime} &=& \varphi ^{\prime} (x_{2}, p^{\prime}_{\perp})=4\left(\frac{\pi}{\beta^{2}}\right)^{\frac{3}{4}} \sqrt{\frac{d p_{z}^{\prime}}{d x_{2}}} \exp(-\frac{p^{\prime 2}_{z}+p^{\prime 2}_{\perp}}{2\beta^{\prime 2}}),
\\
\varphi^{\prime\prime}_{p} &=& \varphi ^{\prime\prime}_{p} (x_{2}, p^{\prime\prime}_{\perp})=\sqrt{\frac{2}{\beta^{2}}}\varphi^{\prime}, \quad \frac{d p^{\prime}_{z}}{d x_{2}}=\frac{e^{\prime}_{1}e_{2}}{x_{1}x_{2}M^{\prime}_{0}}.
\en

Using the formulas provided above and taking the integration rules given in \cite{cheng h y 2004}, one can derive the explicit expressions for the $B\to T$ transition form factors, which are collected in Appendix \ref{FL2}.

\subsection{Observables Calculation Framework}

 Based on the form factors derived in the previous section, we now proceed to calculate the partial decay widths of the semi-leptonic decays $B \to T\ell\nu_\ell$
\be
\frac{d \Gamma}{d q^2} &=& \sum_{i=L,\pm} \frac{d \Gamma_{i}}{d q^{2}}, \label{width}\\
\frac{d \Gamma_{L,\pm}}{d q^2} &=& \frac{|G_{F}V_{CKM}|^{2}\sqrt{\lambda_{T}}}{256 m^{3}_{B}\pi^{3}q^2}\bigg(1-\frac{m^{2}_{\ell}}{q^2}\bigg)^{2}(X_{L},X_{\pm}),
\en
where
$\lambda_{T} = \lambda(m^{2}_{B},m^{2}_{T},q^2) = (m^{2}_{B}-m^{2}_{T}-q^2)^2-4 m^{2}_{T} q^2$, and $X_L, X_\pm$ are defined as
\be
X_L &=& \frac{\lambda_T}{9 m^{2}_{T}m^{2}_{B}}\Big[(2 q^2+m^{2}_{\ell})h^2_0(q^2)+3\lambda_Tm^{2}_{\ell}A^{2}_{0}(q^2)\Big],\\
X_{\pm} &=& \frac{2 q^2}{3}(2 q^2+m^{2}_{\ell})\frac{\lambda_T}{8 m^{2}_{T}m^{2}_{B}}\bigg[(m_{B}+m_T)A_1(q^2)
\mp\frac{\sqrt{\lambda_T}}{m_{B}+m_T}V(q^2)\bigg]^2,\\
h_0(q^2) &=& \frac{1}{2 m_T}\bigg[(m^{2}_{B}-m^{2}_{T}-q^2)(m_{B}+m_T)A_1(q^2)-\frac{\lambda_T}{m_{B}+m_T}A_2(q^2) \bigg].
\en
The subscripts $(L, \pm)$ refer to the three polarization states of the tensor meson along its momentum direction, namely $0$ and $\pm 1$.
Here, $m_\ell$ denotes the lepton mass, and $q^2$ represents the squared momentum transfer to the lepton pair.

In terms of the angular distributions, one can define the partial forward-backward asymmetry $\mathcal{A}_{FB}$ of the lepton,
\be
\frac{d \mathcal{A}_{FB}}{d q^2} &=& \frac{\int^{1}_{0} dz(d \Gamma/d q^2 dz) -\int^{0}_{-1} dz(d \Gamma/d q^2 dz)}{\int^{1}_{0} dz(d \Gamma/d q^2 dz) +\int^{0}_{-1} dz(d \Gamma/d q^2 dz)},
\en
where $z = \cos\theta$, and $\theta$ represents the polar angle of the lepton with respect to the moving direction of the tensor meson in the lepton pair rest frame. The explicit expression for such asymmetry reads
\be
\frac{d \mathcal{A}_{F B}}{d q^2} &=& \frac{1}{X_L+X_++X_-}\bigg(\left.\frac{\lambda_T}{6 m^{2}_{T}m^{2}_{B}}2 m^{2}_{\ell}\sqrt{\lambda_T}h_0(q^2)A_0(q^2) \right.\notag\\&&\left. -\frac{\lambda_T}{8 m^{2}_{T}m^{2}_{B}}4 q^4 \sqrt{\lambda_T}A_1(q^2)V(q^2) \right. \bigg).\label{afb}
\en

Integrating over the $q^2$ from $m^2_\ell$ to $(m_B-m_T)^2$ in Eq. (\ref{width}) and Eq. (\ref{afb}), we can obtain the total decay widths and the integrated $\mathcal{A}_{FB}$ for the semi-leptonic decay $B \to T \ell \nu_\ell$.
Since there are three different polarizations, it is also meaningful to define the polarization fraction
\be
f_L &=& \frac{\Gamma_L}{\Gamma_L+\Gamma_++\Gamma_-}.
\en

\section{Numerical results and discussions} \label{numer}
\begin{table}[H]
\caption{The values of the input parameters \cite{chang-qin:2112,PDG}.}
\label{constant}
\begin{tabular*}{16.5cm}{@{\extracolsep{\fill}}l|cccccc}
  \hline\hline
\textbf{Mass(\text{GeV})} &$m_{b}=4.8$ &$m_{c}=1.4$&$m_{d,u}=0.25$&$m_{s}=0.37$\\[1ex]
&$m_{e}=0.000511$&$m_{\mu}=0.106$&$m_{\tau}=1.78$   \\[1ex]
& $ m_{B}=5.28$  & $m_{B_{s}}=5.37$& $m_{a_2(1320)}=1.32 $& $ m_{f_2(1270)}=1.27$ \\[1ex]
& $m_{D_2^*(2460)}=2.46$ &$ m_{D_{s2}^*(2573)}=2.57$ & $m_{K_2^*(1430)}=1.43$ \\[1ex]
\hline
\end{tabular*}
\begin{tabular*}{16.5cm}{@{\extracolsep{\fill}}l|cccc}
  \hline
  {{\textbf{Shape parameters(GeV)}}} &$\beta_{B}=0.555^{+0.060}_{-0.060}$&$\beta_{B_s}=0.626^{+0.045}_{-0.045}$ \\[1ex]
  &$\beta_{f_2^\prime}=0.348^{+0.006}_{-0.006}$&$\beta_{K_2^*}=0.313^{+0.010}_{-0.010}$ \\[1ex]
  &$\beta_{D_2^*}=0.429^{+0.013}_{-0.013}$&$\beta_{D_{s2}^*}=0.530^{+0.019}_{-0.019}$ \\[1ex]
  &$\beta_{a_2}=\beta_{f_2}=0.312^{+0.006}_{-0.006}$ \\[1ex]
\hline\hline
{{\textbf{CKM}}} &$V_{cb}=(41.1\pm1.2) \times 10^{-3}$&$V_{ub}=(3.82\pm0.2) \times 10^{-3}$& \\[1ex]
\hline\hline
\textbf{Lifetimes(s)}&$\tau_{B}=(1.638\pm0.004)\times 10^{-12}$&$\tau_{B_s}=(1.516\pm0.006)\times 10^{-12}$\\[1ex]
\hline\hline
\end{tabular*}
\end{table}

The input parameters, including the initial and final meson (quark) masses, the CKM matrix elements, the $B_{(s)}$ meson lifetime, and the shape parameters of the relevant mesons, are listed in Table \ref{constant}. Based on the theoretical framework presented in Section \ref{form} and the input parameters provided in Table \ref{constant}, one can calculate the $B \to T$ transition form factors at $q^2 = 0$, which are listed in Table \ref{BtoT} in Appendix \ref{FL} with other theoretical results for comparison.

In the calculations, we adopt the reference frame $q^+ = 0$, where the form factors can only be computed in the spacelike momentum transfer region ($q^2 = -q_\perp^2 \le 0$). However, physical processes occur in the timelike region ($q^2 > 0$), necessitating the analytic continuation of the form factors from the spacelike to the timelike region.
To achieve this, we employ a double-pole parametrization formula to extend the form factors
\be
F(q^2) &=& \frac{F(0)}{1 - a(q^2/m^2_B) + b(q^4/m^4_B)},
\en
where $F(q^2)$ represents the various form factors ($F_1, F_0, V, A_0, A_1, A_2$), $F(0)$ is the value of the form factor at $q^2 = 0$. The fitting parameters $a$ and $b$ can be obtained by performing a three-parameter fit to the form factors in the spacelike region $-15 \, \text{GeV}^2 \leq q^2 \leq 0$. The uncertainties arise from the decay constants of the initial and final state mesons.

Based on the these form factors, we can calculate the physical observables for the relevant semi-leptonic decays, including the branching ratios, polarization fractions $f_L$, and forward-backward asymmetries $\mathcal{A}_{FB}$. The results are presented in Tables \ref{tabD2}, \ref{tab2}, \ref{tab1x} and \ref{tab2x}, respectively. For the branching ratios, some comments are in order.

\begin{itemize}
\item
For the decays $B\to (a_2, f_2, K^{*}_2)\ell\nu_\ell$ induced by the $b\to u$ transition, their branching ratios lie in the range $10^{-5}\sim10^{-4}$. While the branching ratios of the decays $B_{(s)}\to D^{*}_{(s)2}\ell\nu_\ell$ can reach $10^{-4}\sim10^{-3}$ because the related CKM matrix element $V_{cb}$ is about one order larger than $V_{ub}$.
\item
Due to phase space constraints, the branching ratios of the decays $B \to (a_2, f_2, K^{*}_2)\tau\nu_\tau$ are estimated to be one-fifth to one-fourth of those of the corresponding decays $B \to (a_2, f_2, K^{*}_2)\ell^\prime\nu_{\ell^\prime}$. As to the branching ratios of the decays $B_{(s)}\to D^{*}_{(s)2}\tau\nu_\tau$ and $B_{(s)}\to D^{*}_{(s)2}\ell^{\prime}\nu_{\ell^\prime}$, the difference is more significant, the former are almost one-twentieth of the latter. This is because the $\tau$ lepton mass is comparable to the mass of $D^{*}_{(s)2}$ meson, leading to severe compression from the phase space.
\item
In Table \ref{tab2}, the branching ratios of most considered decays are consistent with the results from other approaches, such as the PQCD \cite{wang-wei:1008}, the QCD sum rules (QCDSRs) \cite{R.Khosravi:1503}, the light-cone sum rules (LCSRs) \cite{zuo 21} and the relativistic quark model (RQM) \cite{Faustov 14,Ebert 11}. While the branching ratios of the decays
$B \to f_2\ell^\prime\nu_{\ell^\prime}$ are larger than other theoretical calculations as shown in Table \ref{tab2}, which need further
clarification. Anyway, all the branching ratios of these considered decays are larger than $10^{-5}$, indicating promising prospects for measuring these processes in ongoing Belle II and LHCb experiments.
\end{itemize}

\begin{table}[H]
\caption{The branching ratios $(10^{-5})$ of the semi-leptonic decays $B \to (a_2, f_2, K^{*}_2) \ell \nu_{\ell}$.}
\setlength{\tabcolsep}{0.3mm}{
\begin{center}
\begin{tabular}{c|c|ccccccc}
\hline\hline
 Reference&This work &\quad PQCD \cite{wang-wei:1008}\;\;\quad&\quad QCDSR \cite{R.Khosravi:1503}\quad &\quad LCSR \cite{zuo 21}\quad&\quad RQM \cite{Faustov 14,Ebert 11}\quad\\
 \hline
$B^{+}\to \ a_{2}^{0} e^{+}\nu_{e}$&$\quad 9.25^{+0.02+1.11+1.67}_{-0.02-2.91-0.59} \quad$&$-$&$-$&$5.60$&$-$\\
$B^{+}\to \ a_{2}^{0} \mu^{+}\nu_{\mu}$&$9.22^{+0.02+1.11+1.66}_{-0.02-2.90-0.57}$&$-$&$-$&$5.50$&$-$\\
$B^{+}\to \ a_{2}^{0} \tau^{+}\nu_{\tau}$&$2.15^{+0.01+0.30+0.34}_{-0.01-0.61-0.36}$&$-$&$-$&$1.10$&$-$\\
\hline
$B^{0}\to \ a_{2}^{-} e^{+}\nu_{e}$&$8.56^{+0.02+1.03+1.55}_{-0.02-2.70-0.55}$&$11.60$&$8.20$&$10.30$&$12.70$\\
$B^{0}\to \ a_{2}^{-} \mu^{+}\nu_{\mu}$&$8.54^{+0.02+1.03+1.54}_{-0.02-2.69-0.53}$&$11.60$&$8.20$&$10.20$&$12.70$\\
$B^{0}\to \ a_{2}^{-} \tau^{+}\nu_{\tau}$&$2.00^{+0.01+0.28+0.31}_{-0.01-0.56-0.33}$&$4.10$&$5.10$&$2.00$&$2.80$\\
\hline
$B^{+}\to \ f_{2}^{0} e^{+}\nu_{e}$&$12.44^{+0.03+3.10+2.20}_{-0.03-4.70-1.05}$&$6.90$&$7.70$&$6.00$&$-$\\
$B^{+}\to \ f_{2}^{0} \mu^{+}\nu_{\mu}$&$12.38^{+0.03+3.08+2.19}_{-0.03-4.67-1.05}$&$6.90$&$7.70$&$6.00$&$-$\\
$B^{+}\to \ f_{2}^{0} \tau^{+}\nu_{\tau}$&$2.88^{+0.01+0.66+0.45}_{-0.01-0.93-0.24}$&$2.50$&$5.30$&$1.20$&$-$\\
\hline
$B^{0}\to \ f_{2}^{-} e^{+}\nu_{e}$&$11.54^{+0.03+2.89+2.02}_{-0.03-4.34-0.99}$&$-$&$-$&$-$&$-$\\
$B^{0}\to \ f_{2}^{-} \mu^{+}\nu_{\mu}$&$11.47^{+0.03+2.85+2.03}_{-0.03-4.33-0.97}$&$-$&$-$&$-$&$-$\\
$B^{0}\to \ f_{2}^{-} \tau^{+}\nu_{\tau}$&$2.67^{+0.01+0.62+0.42}_{-0.01-0.86-0.23}$&$-$&$-$&$-$&$-$\\
\hline\hline
$B_{s}^{0}\to \ K_{2}^{*-} e^{+}\nu_{e}$&$7.90^{+0.03+0.90+2.45}_{-0.03-1.29-0.82}$&$7.30$&$6.50$&$9.60$&$13.30$\\
$B_{s}^{0}\to \ K_{2}^{*-} \mu^{+}\nu_{\mu}$&$7.87^{+0.03+0.90+2.43}_{-0.03-1.27-0.82}$&$7.30$&$6.50$&$9.50$&$13.30$\\
$B_{s}^{0}\to \ K_{2}^{*-} \tau^{+}\nu_{\tau}$&$1.77^{+0.01+0.23+0.46}_{-0.01-0.23-0.21}$&$2.50$&$3.50$&$1.70$&$3.60$\\
\hline\hline
\end{tabular}\label{tab2}
\end{center}}
\end{table}
In what follows, we address the issue of polarization characteristics for our considered decays. For the decays $B\to T\ell^\prime\nu_{\ell^\prime}$, their longitudinal polarization fractions $f_L$ are close to $70\%$, while those of the decays $B\to T\tau\nu_{\tau}$ are a little smaller. The least values are from the decays $B\to D_{(s)2}^*\tau\nu_\tau$ and about $56\%$. Even so, all these decays are dominated by longitudinal polarization. This feature is also supported by other theoretical calculations. The difference in longitudinal polarization fractions between the decays $B\to (a_2, f_2, K^{*}_2)\ell^\prime\nu_{\ell^\prime}$ and $B\to (a_2, f_2, K^{*}_2)\tau\nu_{\tau}$ is approximately $(6\sim8)\%$, while that from the decays $B_{(s)}\to D_{(s)2}^*{\ell^\prime}\nu_{\ell^\prime}$ and $B_{(s)}\to D_{(s)2}^*\tau\nu_{\tau}$ is more significant and can arrive at $15\%$.

\begin{table}[H]
\caption{The branching ratios $(10^{-3})$ of the semi-leptonic decays $B_{(s)} \to D^*_{(s)2}\ell \nu_{\ell}$.}
\setlength{\tabcolsep}{0.3mm}{
\begin{center}
\begin{tabular}{c|c|cccccccc}
\hline\hline
 Reference&This work &\quad RQM \cite{Faustov 14}\;\;\quad&\quad BT \cite{Mor 97,Ebert 20}\quad&\quad GI \cite{Mor 97}\quad &\quad ISGW \cite{Mor 97}\quad&\quad OPAL \cite{Ake 95}\quad\\
 \hline
$B^{+}\to \ D_{2}^{*0} e^{+}\nu_{e}$&$7.08^{+0.02+1.17+0.03}_{-0.02-1.26-0.48}$&$-$&$4.50\sim8.00$&$7.00$&$7.70$&$8.80$\\
$B^{+}\to \ D_{2}^{*0} \mu^{+}\nu_{\mu}$&$7.02^{+0.02+1.16+0.03}_{-0.02-1.25-0.47}$&$-$&$4.50\sim8.00$&$7.00$&$7.70$&$8.80$\\
$B^{+}\to \ D_{2}^{*0} \tau^{+}\nu_{\tau}$&$0.42^{+0.00+0.05+0.06}_{-0.00-0.02-0.01}$&$-$&$-$&$-$&$-$&$-$\\
\hline
$B^{0}\to \ D_{2}^{*-} e^{+}\nu_{e}$&$6.56^{+0.02+1.08+0.02}_{-0.02-1.17-0.44}$&$-$&$4.50\sim8.00$&$-$&$-$&$-$\\
$B^{0}\to \ D_{2}^{*-} \mu^{+}\nu_{\mu}$&$6.50^{+0.02+1.07+0.03}_{-0.02-1.16-0.44}$&$-$&$4.50\sim8.00$&$-$&$-$&$-$\\
$B^{0}\to \ D_{2}^{*-} \tau^{+}\nu_{\tau}$&$0.34^{+0.00+0.06+0.01}_{-0.00-0.06-0.02}$&$-$&$-$&$-$&$-$&$-$\\
\hline\hline
$B_{s}^{0}\to \ D_{s2}^{*-} e^{+}\nu_{e}$&$8.20^{+0.03+0.37+0.47}_{-0.03-0.46-0.28}$&$6.70$&$7.5$&$-$&$-$&$-$\\
$B_{s}^{0}\to \ D_{s2}^{*-} \mu^{+}\nu_{\mu}$&$8.20^{+0.03+0.98+0.43}_{-0.06-0.42-0.31}$&$6.70$&$-$&$-$&$-$&$-$\\
$B_{s}^{0}\to \ D_{s2}^{*-} \tau^{+}\nu_{\tau}$&$0.42^{+0.00+0.05+0.06}_{-0.00-0.03-0.01}$&$0.29$&$-$&$-$&$-$&$-$\\
\hline\hline
\end{tabular}\label{tabD2}
\end{center}}
\end{table}

\begin{table}[H]
\caption{The longitudinal polarization fractions $f_{L}$
(\%) for the semi-leptonic decays $B\to T\ell\nu_{\ell}$.}
\begin{center}
\scalebox{0.9}{
\begin{tabular}{|c|c|c|c|}
\hline\hline
Decay modes&$B^{0}\to \ a_{2}^{-} e^{+}\nu_{e}$&$B^{0}\to \ a_{2}^{-}  \mu^{+}\nu_{\mu}$&$B^{0}\to \ a_{2}^{-}  \tau^{+}\nu_{\tau}$\\
 \hline\hline
This work&$69.53^{+0.17+8.19+15.29}_{-0.17-25.27-6.98}$&$69.56^{+0.17+8.33+15.26}_{-0.17-25.23-6.81}$
&$62.47^{+0.15+9.49+11.92}_{-0.15-20.22-2.16}$\\
PQCD \cite{wang-wei:1008}&$73.3$&$73.3$&$69.9$\\
LCSR \cite{zuo 21}&$71.8$&$71.8$&$59.6$\\
\hline\hline
Decay modes&$B^{0}\to \ f_{2}^{-} e^{+}\nu_{e}$&$B^{0}\to \ f_{2}^{-}  \mu^{+}\nu_{\mu}$&$B^{0}\to \ f_{2}^{-}  \tau^{+}\nu_{\tau}$\\
 \hline\hline
This work&$72.58^{+0.17+8.19+15.29}_{-0.17-25.27-6.98}$&$72.58^{+0.17+8.33+15.26}_{-0.17-25.23-6.81}$
&$64.74^{+0.15+9.49+11.92}_{-0.15-20.22-2.16}$\\
PQCD \cite{wang-wei:1008}&$74.9$&$74.9$&$71.9$\\
LCSR \cite{zuo 21}&$68.9$&$68.8$&$56.7$\\
\hline\hline
 Decay modes&$B^{0}\to \ D_{2}^{*-} e^{+}\nu_{e}$&$B^{0}\to \ D_{2}^{*-} \mu^{+}\nu_{\mu}$&$B^{0}\to \ D_{2}^{*-}  \tau^{+}\nu_{\tau}$\\
 \hline\hline
This work&$70.65^{+0.17+11.89+0.33}_{-0.17-12.94-4.93}$&$70.60^{+0.17+11.88+0.36}_{-0.17-12.93-4.91}$
&$55.98^{+0.14+9.37+0.63}_{-0.14-10.10-3.64}$\\
\hline\hline
Decay modes&$B_{s}^{0}\to \ K_{2}^{*-} e^{+}\nu_{e}$&$B_{s}^{0}\to \ K_{2}^{*-} \mu^{+}\nu_{\mu}$
&$B_{s}^{0}\to \ K_{2}^{*-} \tau^{+}\nu_{\tau}$\\
 \hline\hline
This work&$67.99^{+0.27+7.54+21.22}_{-0.27-13.83-6.55}$&$68.08^{+0.27+7.67+21.13}_{-0.27-13.74-6.62}$
&$62.75^{+0.25+8.72+17.03}_{-0.25-9.78-7.18}$\\
PQCD \cite{wang-wei:1008}&$68.3$&$68.3$&$64.1$\\
LCSR \cite{zuo 21}&$77.6$&$77.5$&$65.8$\\
\hline\hline
Decay modes&${B_{s}^{0}}\to \ D_{s2}^{*-} e^{+}\nu_{e}$&${B_{s}^{0}}\to \ D_{s2}^{*-} \mu^{+}\nu_{\mu}$&${B_{s}^{0}}\to \ D_{s2}^{*-}  \tau^{+}\nu_{\tau}$\\
 \hline\hline
This work&$71.52^{+0.28+8.34+4.92}_{-0.28-5.12-2.21}$&$71.21^{+0.28+8.31+4.86}_{-0.28-5.06-2.22}$
&$56.28^{+0.22+6.39+3.21}_{-0.22-3.51-1.54}$\\
\hline\hline
\end{tabular}\label{tab1x}}
\end{center}
\end{table}

\begin{figure}[H]
\vspace{0.32cm}
  \centering
  \subfigure[]{\includegraphics[width=0.30\textwidth]{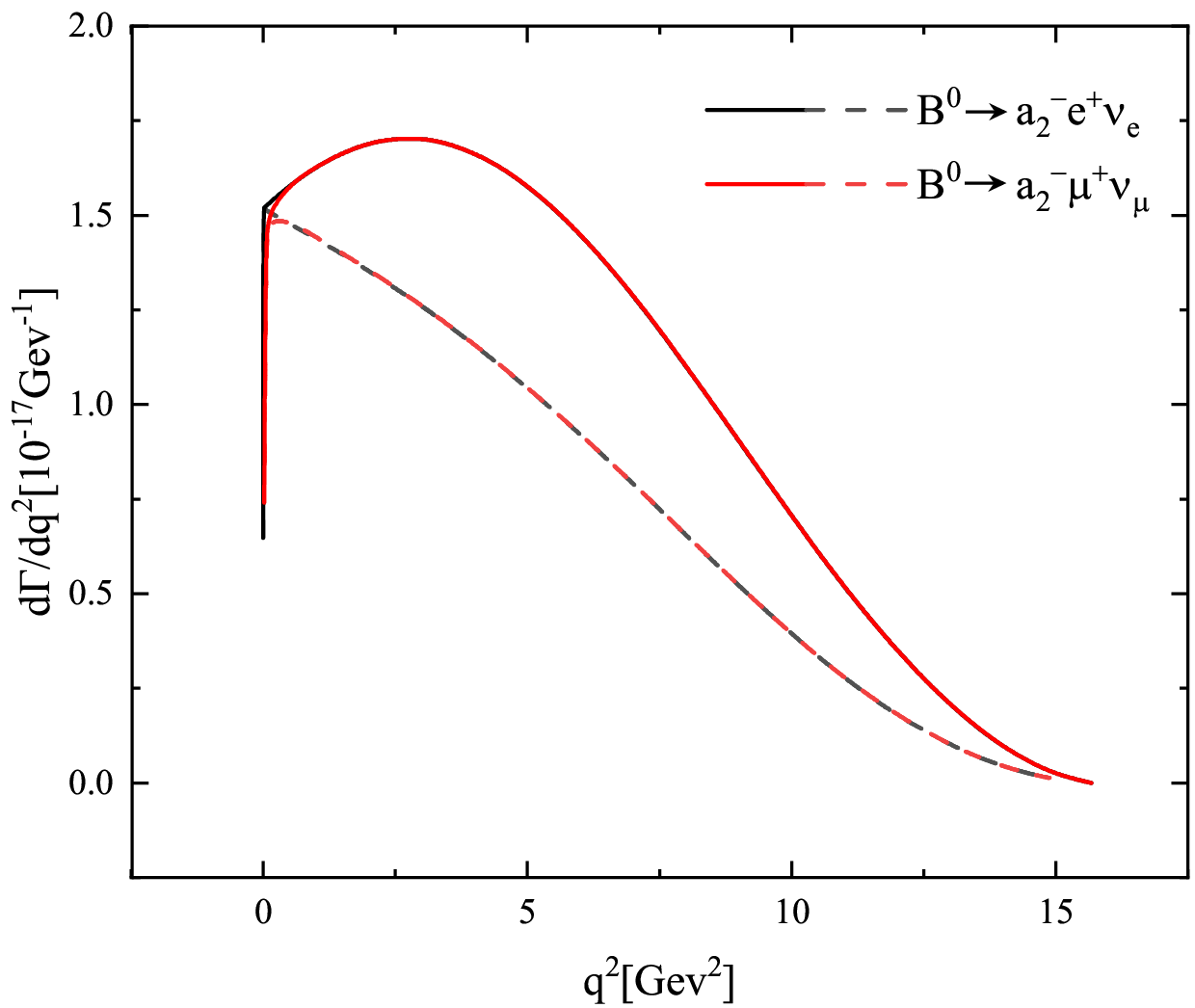}\quad}
  \subfigure[]{\includegraphics[width=0.30\textwidth]{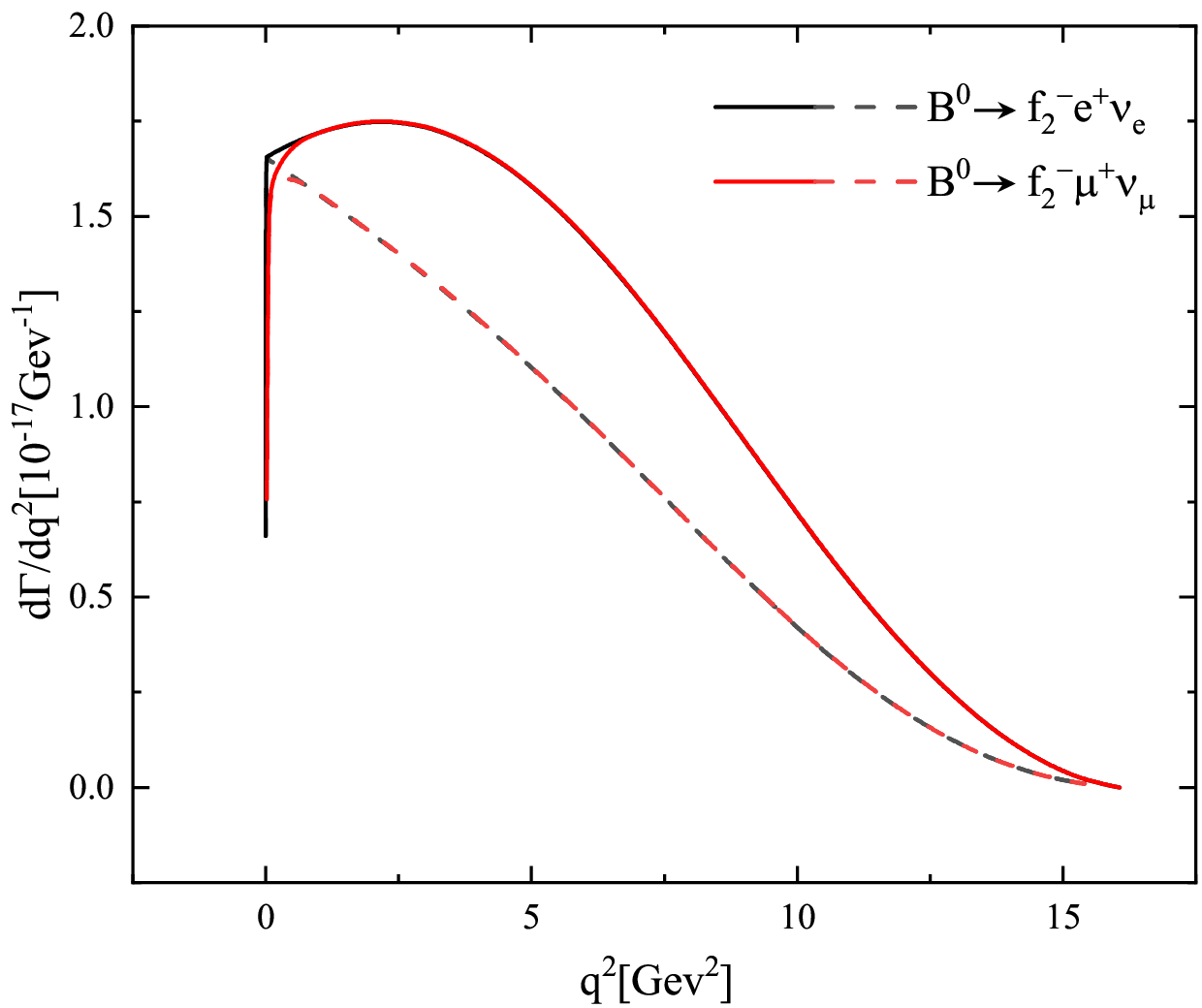}\quad}
  \subfigure[]{\includegraphics[width=0.30\textwidth]{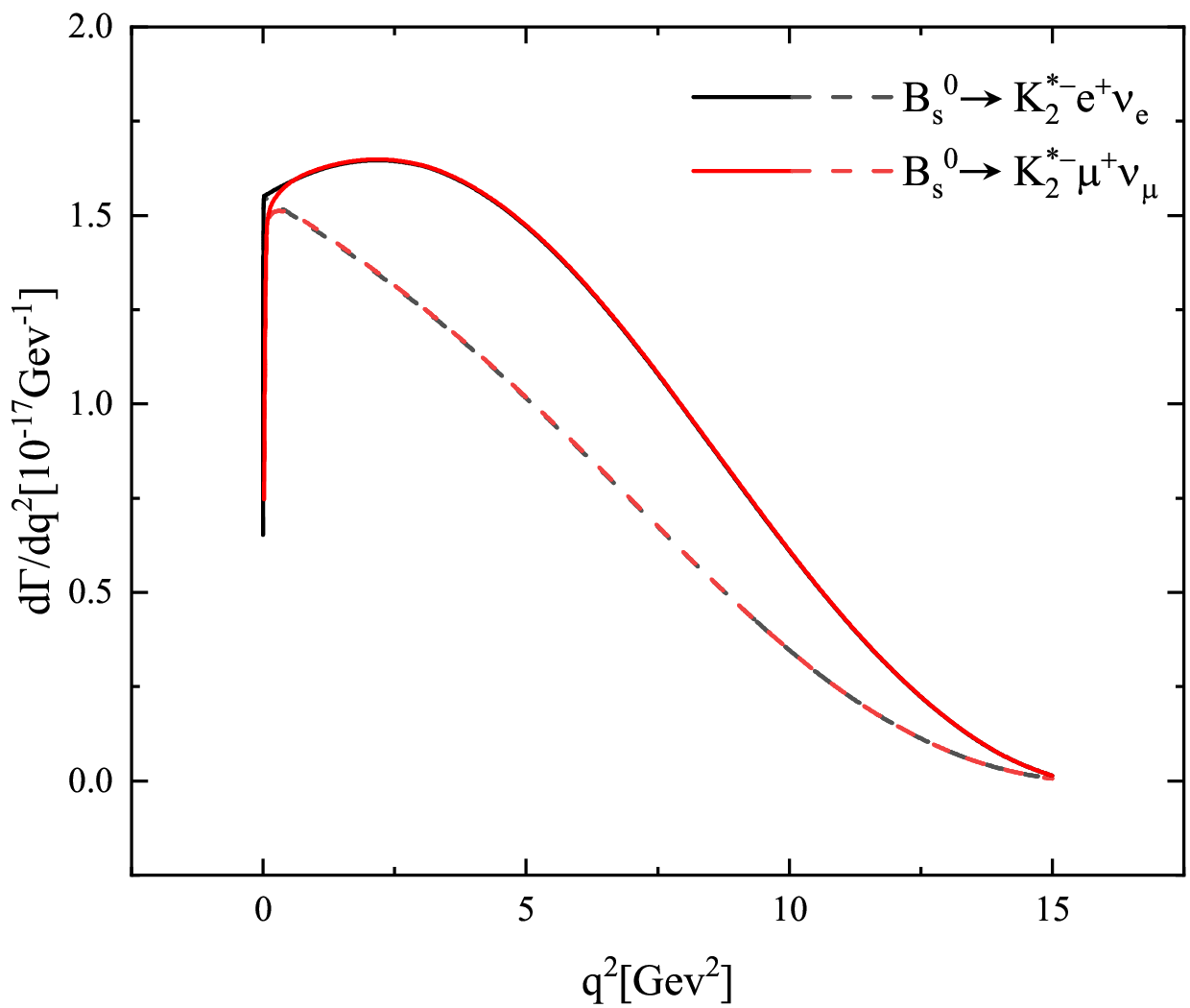}}\\
  \subfigure[]{\includegraphics[width=0.30\textwidth]{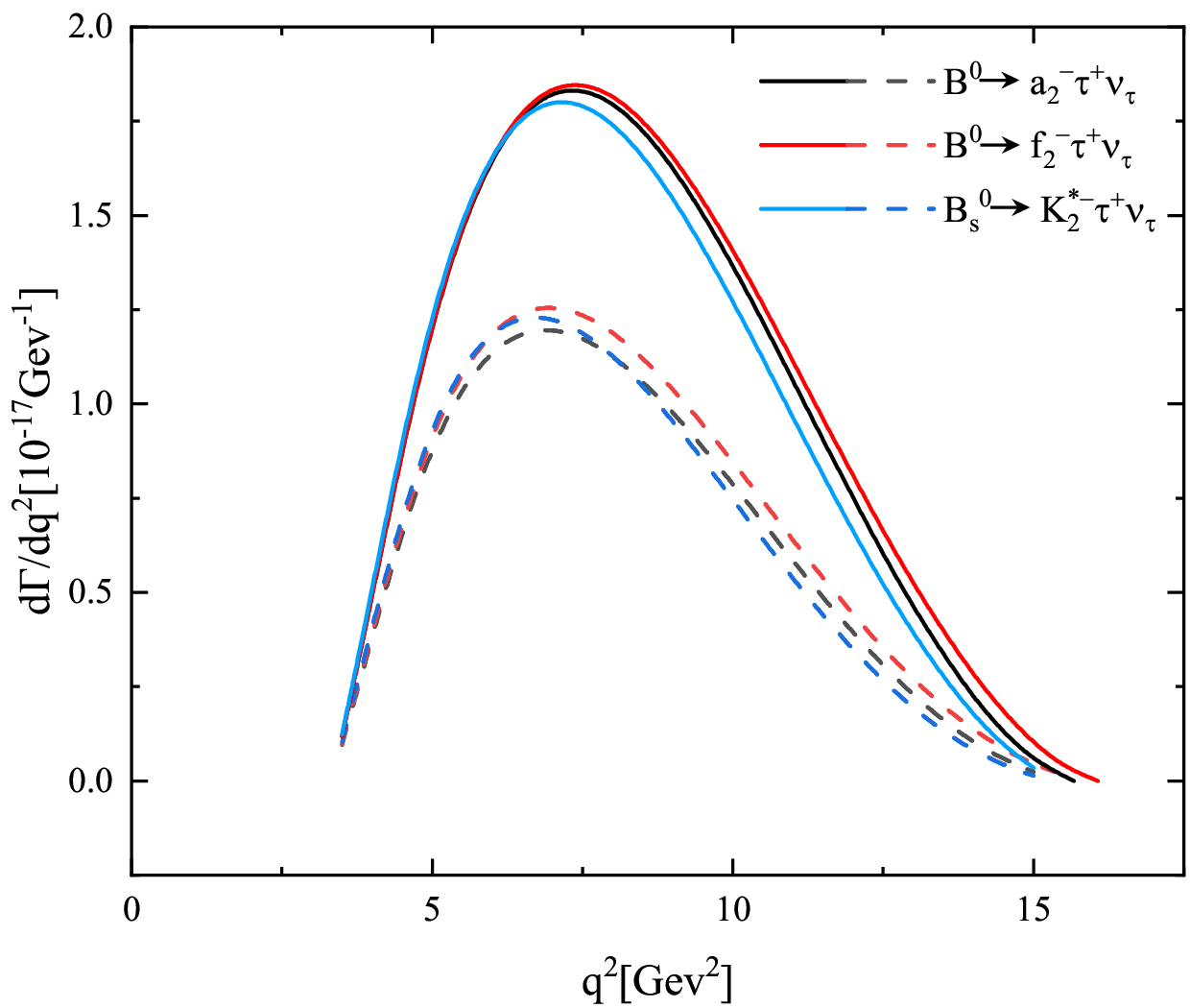}\quad}
  \subfigure[]{\includegraphics[width=0.30\textwidth]{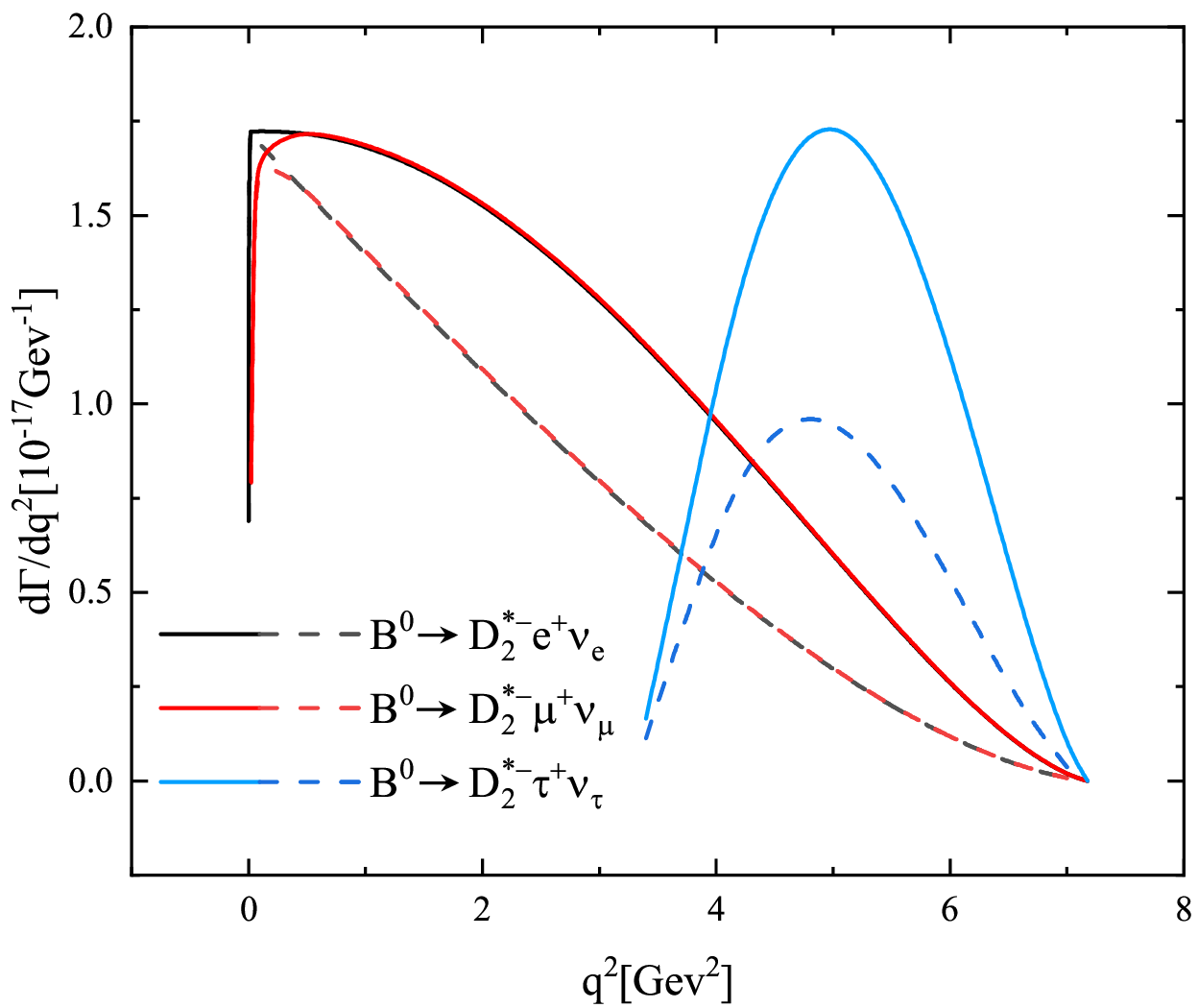}\quad}
  \subfigure[]{\includegraphics[width=0.30\textwidth]{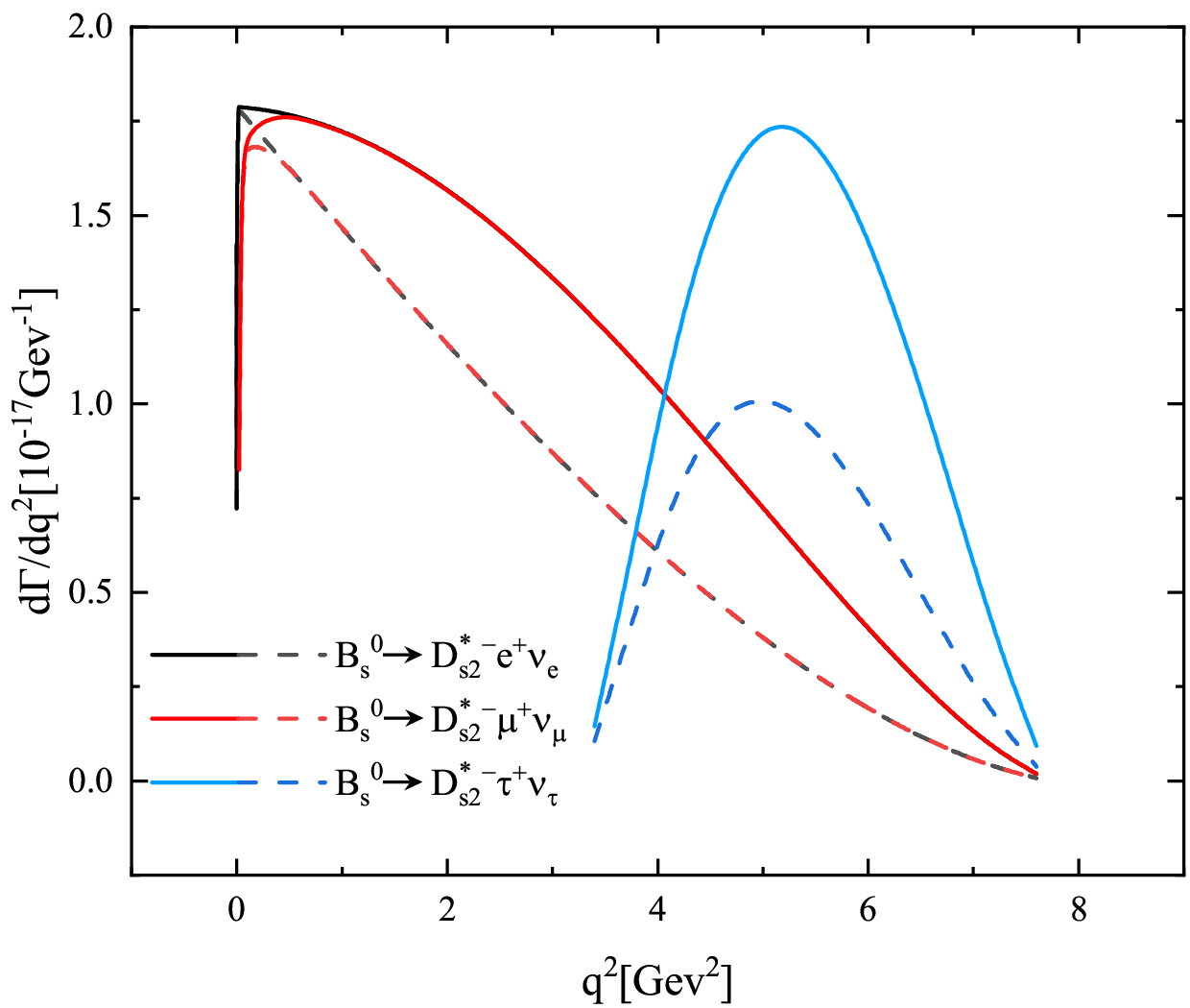}}
  \caption{The $q^2$ dependence of the differential decay widths $d\Gamma/dq^2$ (solid lines) and $d\Gamma_L/dq^2$ (dashed
lines) for the decays $B\to a_2 \ell^\prime\nu_{\ell^\prime}$ (a), $B\to f_2 \ell^\prime\nu_{\ell^\prime}$ (b), $B_s\to K^*_2 \ell^\prime\nu_{\ell^\prime}$ (c), $B_{(s)}\to (a_2,f_2,K^*_2) \tau\nu_{\tau}$ (d), $B\to D^*_2 \ell\nu_{\ell}$ (e) and $B_s\to D^*_{s2} \ell\nu_{\ell}$ (f).}\label{polari}
\end{figure}

The dependence of the differential decay width $d\Gamma_{(L)}/dq^2$ with $L$ representing the contribution from the longitudinal polarization on $q^2$ for each decay is shown in Figure \ref{polari}. It can be observed that the values of $d\Gamma_{(L)}/dq^2$ are zero at the zero-recoil point ($q^2 = q^2_{\text{max}}$). Furthermore, the polarization characteristics for the decays $B\to T\ell^\prime\nu_{\ell^\prime}$ show a clear difference from those of the decays $B\to T\tau\nu_{\tau}$. The transverse polarization components are slightly significant in the decays $B\to T\tau\nu_{\tau}$ compared with those in the decays $B \to T\ell^\prime\nu_{\ell^\prime}$.

%Unlike $f_L$, which primarily reflects the helicity structure of the hadronic matrix elements, $\mathcal{A}_{FB}$ is more sensitive to the relative magnitudes and signs of the form factors.
From Table \ref{tab2x}, we can find that for all the considered decays $B \to T\ell^\prime\nu_{\ell^\prime}$, the forward-backward asymmetries $\mathcal{A}_{FB}$ are negative, which are consistent with the theoretical expectation. If we neglect the $e$ and $\mu$ lepton masses, only the second term in Eq. (\ref{afb}) contributes, leading to negative results. In contrast, for the decays $B \to T\tau\nu_\tau$, the forward-backward asymmetries $\mathcal{A}_{FB}$ are positive. Furthermore, the $\mathcal{A}_{FB}$ values of the decays $B \to (a_2,f_2,K^*_2)\tau\nu_\tau$ are very small, because the large $\tau$ lepton mass makes the term proportional to $m_\ell^2$ shown in Eq. (\ref{afb}) significant, and the two terms cancel each other, resulting in a net value close to zero. However, for the decays $B_{(s)} \to D_{(s)2}^* \tau \nu_\tau$, their $\mathcal{A}_{FB}$ values exhibit relatively large positive values, which is further primarily influenced by the $D_{(s)2}^*$ meson mass.
For the decays $B \to T\ell^\prime\nu_{\ell^\prime}$, our predictions of $\mathcal{A}_{FB}$ are consistent with the results from the PQCD \cite{wang-wei:1008} and non-relativistic quark model (NRQM) \cite{Albertus 14} in both signs and amplitudes. However, the deviations from those of the decays $B \to T\tau\nu_\tau$ are significant, which can be clarified by future LHCb and Belle II experiments.

 \begin{table}[H]
\caption{The forward-backward asymmetries $\mathcal{A}_{FB}$ for the decays $B\to T\ell\nu_{\ell}$ .}
\begin{center}
\scalebox{1.0}{
\begin{tabular}{|c|c|c|c|}
\hline\hline
Decay modes&$B^{0}\to \ a_{2}^{-} e^{+}\nu_{e}$&$B^{0}\to \ a_{2}^{-}  \mu^{+}\nu_{\mu}$&$B^{0}\to \ a_{2}^{-}  \tau^{+}\nu_{\tau}$\\
 \hline\hline
This work&$-0.204$&$-0.202$&$0.012$\\
PQCD \cite{wang-wei:1008}&$-0.186$&$-0.202$&$0.031$\\
\hline\hline
Decay modes&$B^{0}\to \ f_{2}^{-} e^{+}\nu_{e}$&$B^{0}\to \ f_{2}^{-}  \mu^{+}\nu_{\mu}$&$B^{0}\to \ f_{2}^{-}  \tau^{+}\nu_{\tau}$\\
 \hline\hline
This work&$-0.180$&$-0.178$&$0.027$\\
PQCD \cite{wang-wei:1008}&$-0.175$&$-0.175$&$0.048$\\
\hline\hline
 Decay modes&$B^{0}\to \ D_{2}^{*-} e^{+}\nu_{e}$&$B^{0}\to \ D_{2}^{*-} \mu^{+}\nu_{\mu}$&$B^{0}\to \ D_{2}^{*-}  \tau^{+}\nu_{\tau}$\\
 \hline\hline
This work&$-0.146$&$-0.143$&$0.117$\\
\hline\hline
Decay modes&$B_{s}^{0}\to \ K_{2}^{*-} e^{+}\nu_{e}$&$B_{s}^{0}\to \ K_{2}^{*-} \mu^{+}\nu_{\mu}$
&$B_{s}^{0}\to \ K_{2}^{*-} \tau^{+}\nu_{\tau}$\\
 \hline\hline
This work&$-0.183$&$-0.182$&$0.045$\\
PQCD \cite{wang-wei:1008}&$-0.221$&$-0.221$&$-0.024$\\
\hline\hline
Decay modes&${B_{s}^{0}}\to \ D_{s2}^{*-} e^{+}\nu_{e}$&${B_{s}^{0}}\to \ D_{s2}^{*-} \mu^{+}\nu_{\mu}$
&${B_{s}^{0}}\to \ D_{s2}^{*-}  \tau^{+}\nu_{\tau}$\\
 \hline\hline
This work&$-0.139$&$-0.135$&$0.126$\\
NRQM \cite{Albertus 14}&$-0.14$&$-0.12$&$0.06$\\
\hline\hline
\end{tabular}\label{tab2x}}
\end{center}
\end{table}

\begin{figure}[H]
\vspace{0.32cm}
  \centering
  \subfigure[]{\includegraphics[width=0.30\textwidth]{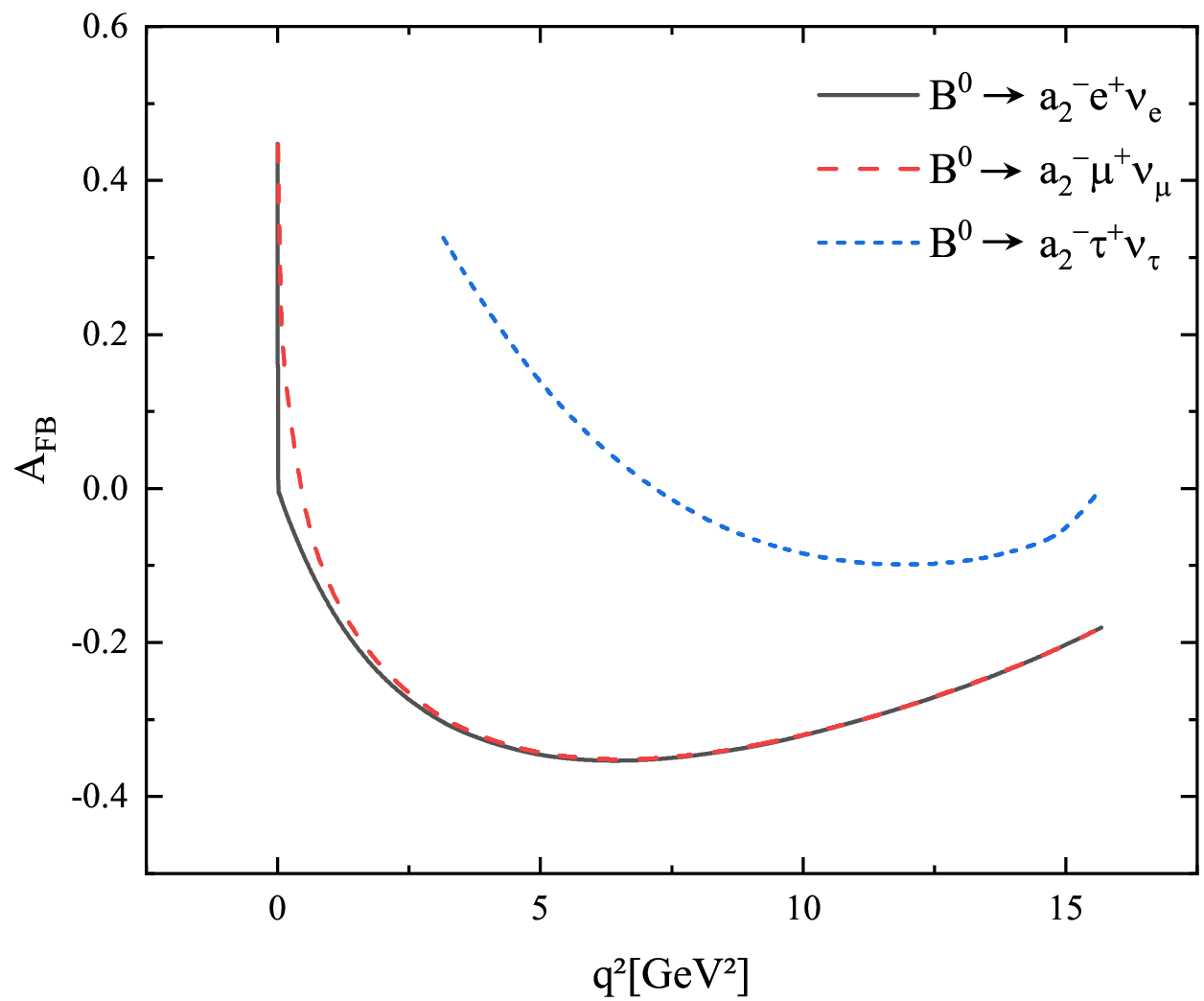}\quad}
  \subfigure[]{\includegraphics[width=0.30\textwidth]{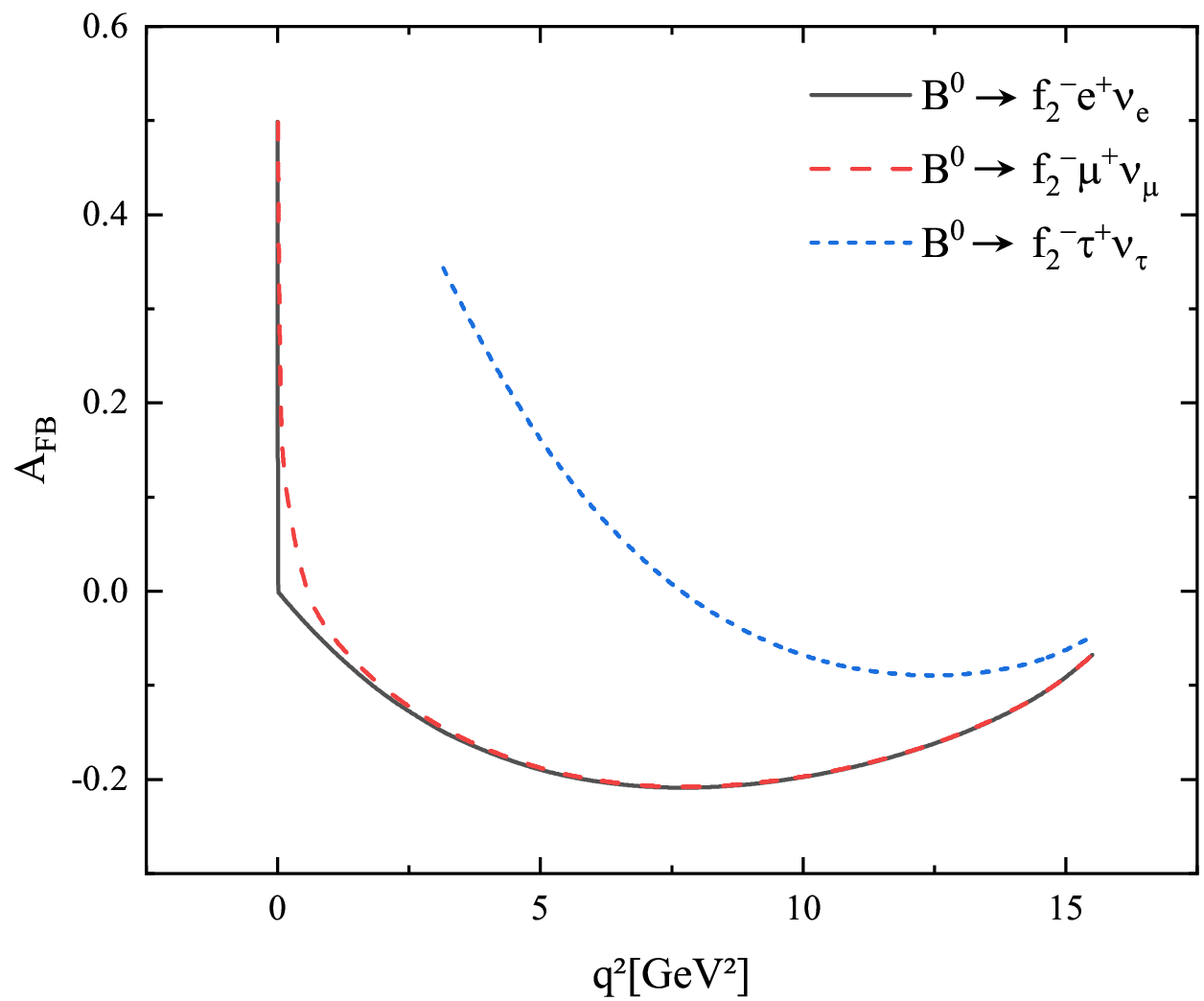}\quad}
  \subfigure[]{\includegraphics[width=0.30\textwidth]{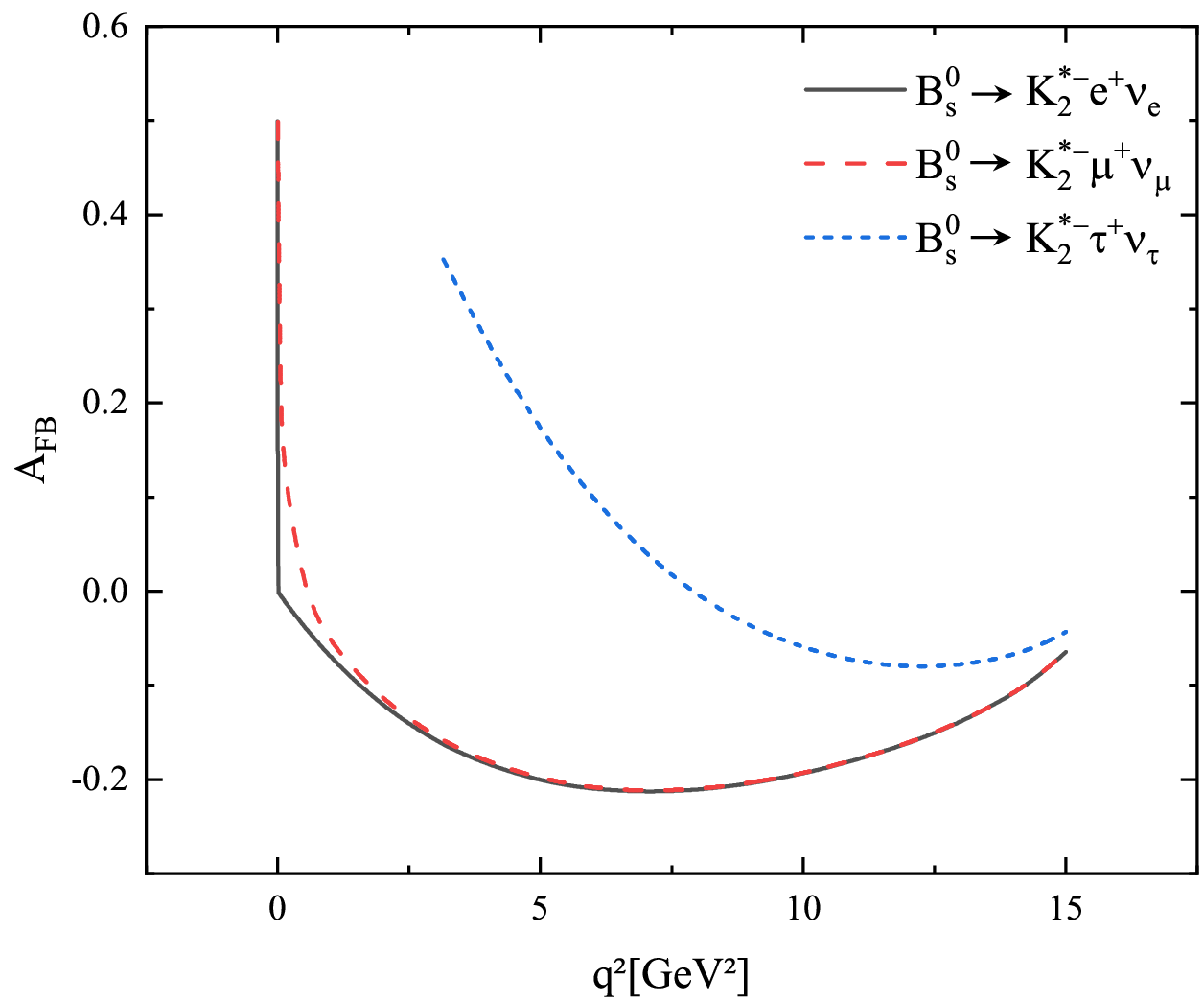}}\\
  \subfigure[]{\includegraphics[width=0.30\textwidth]{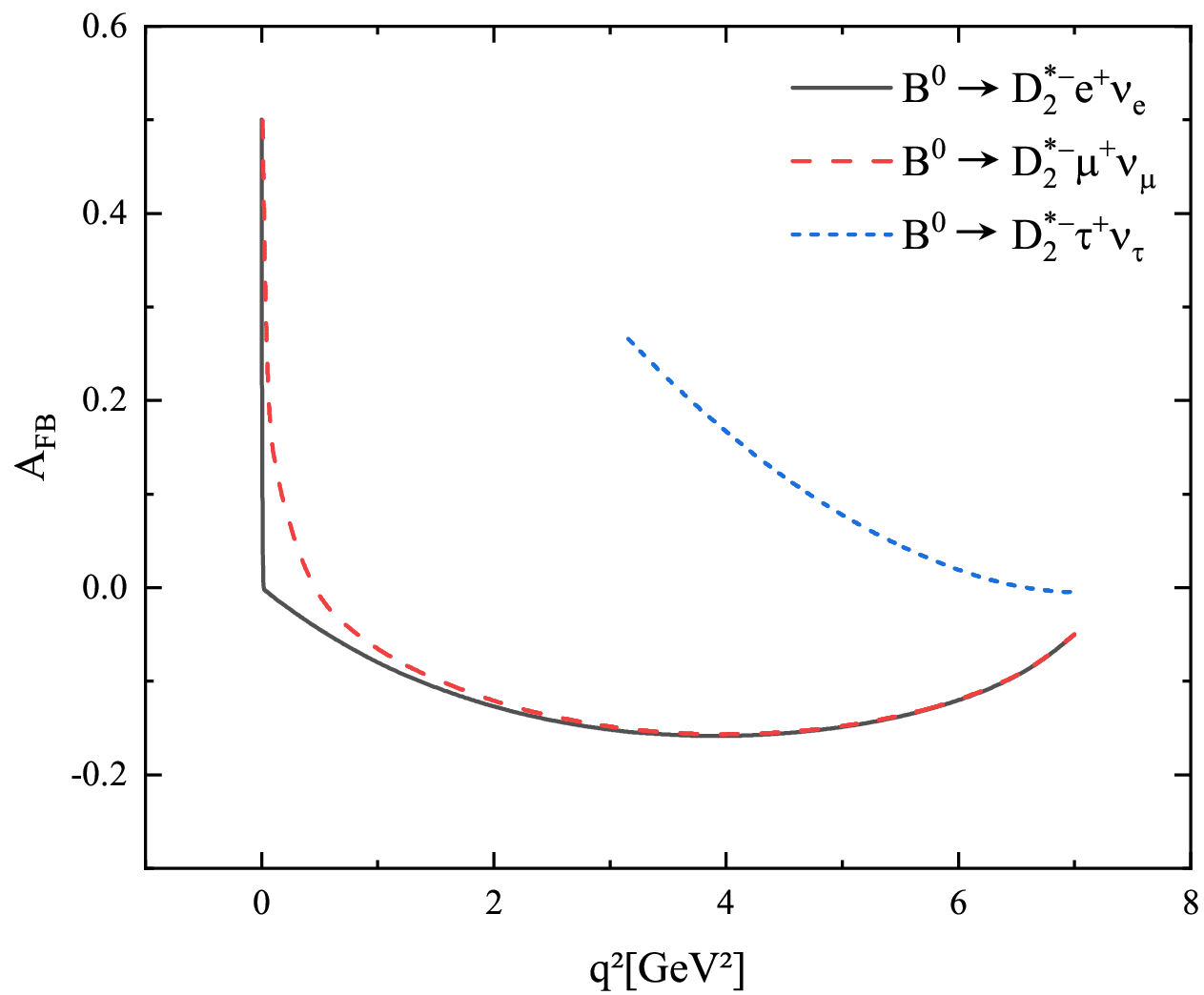}\quad}
  \subfigure[]{\includegraphics[width=0.30\textwidth]{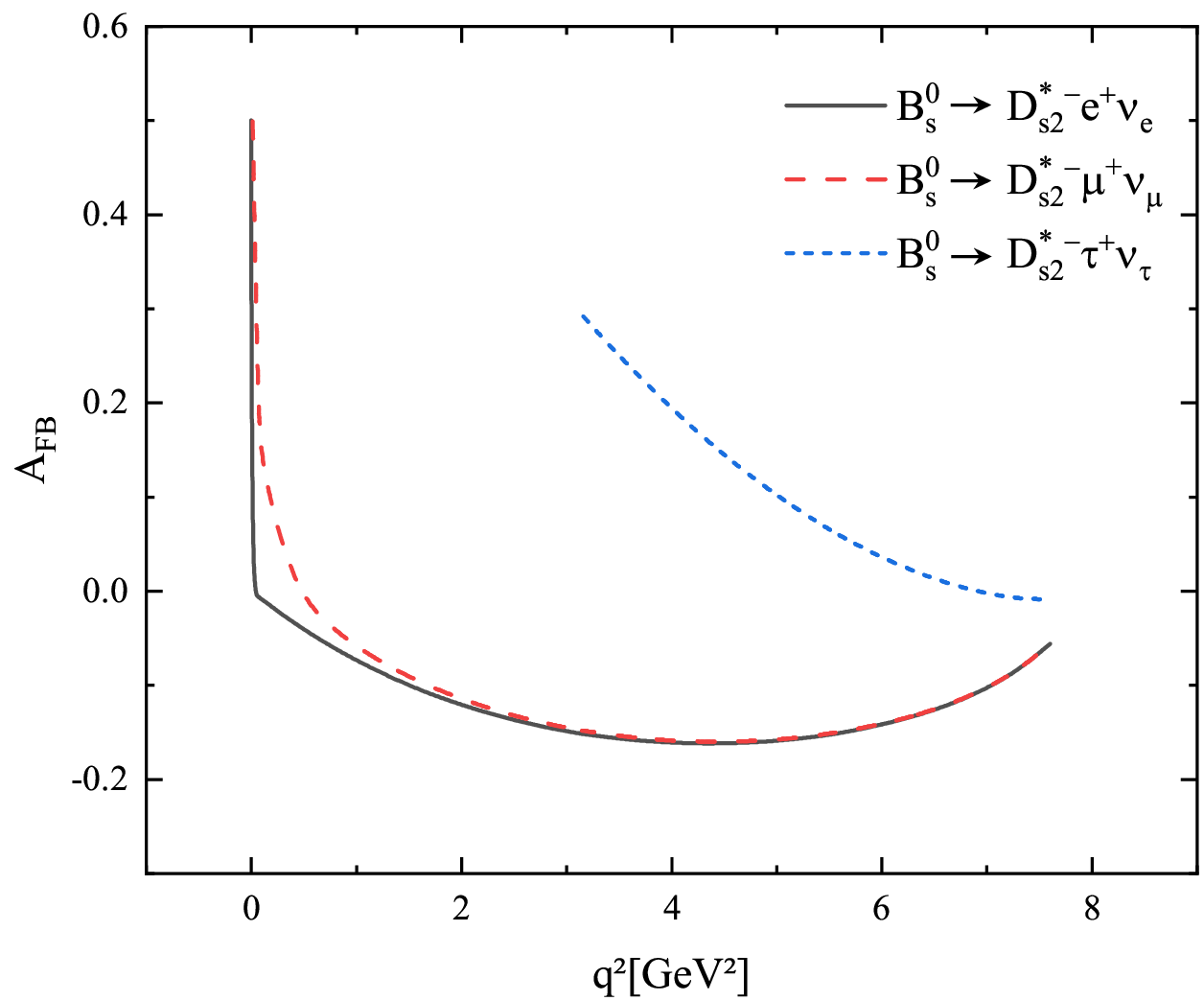}}
  \caption{The $q^{2}$ dependence of the forward-backward asymmetries $\mathcal{A}_{FB}$ for the decays $B\to a_2 \ell\nu_{\ell}$ (a), $B\to f_2 \ell\nu_{\ell}$ (b), $B_s\to K^*_2 \ell\nu_{\ell}$ (c), $B\to D^*_2 \ell\nu_{\ell}$ (d) and $B_s\to D^*_{s2} \ell\nu_{\ell}$ (e).}\label{fig:T1}
\end{figure}

The $q^2$ depedence of the $\mathcal{A}_{FB}$ for different decays $B \to T \ell\nu_\ell$ are shown in Figure \ref{fig:T1}, where the $\mathcal{A}_{FB}$ values exhibit a strong dependence on the lepton mass, mediating by the mass of final state tensor meson as well. It can be clearly seen that the first zero-crossing points increase significantly with the charged lepton mass for each decay.  For the decays $B\to T\ell^\prime\nu_{\ell^\prime}$, their $\mathcal{A}_{FB}$ values are almost entirely negative across the entire $q^2$ range, however, the signs of $\mathcal{A}_{FB}$ for the decays $B\to (a_2,f_2,K^*_2)\tau\nu_{\tau}$ can be either positive or negative within the total $q^2$ range. As to the $\mathcal{A}_{FB}$ values of the decays $B_{(s)}\to (D^*_2,D^*_{s2})\tau\nu_{\tau}$ are entirely positive.

\section{Summary}\label{sum}
Using the the $B \to T$ transition form factors obtained within the framework of the CLFQM, we provide a detailed investigation of the semi-leptonic decays $B \to T\ell\nu_\ell$ . Due to phase space constraints, the branching ratios of the decays $B \to (a_2,f_2,K^*_2) \tau \nu_\tau$ are about $1/5$ to $1/4$ of those of the corresponding decays $B \to (a_2,f_2,K^*_2) \ell^\prime \nu_{\ell^\prime}$. The difference in the branching ratios between the decays $B_{(s)}\to D^{*}_{(s)2}\tau\nu_\tau$ and $B_{(s)}\to D^{*}_{(s)2}\ell^{\prime}\nu_{\ell^\prime}$ is more significant, the former are only $1/20$ of the latter. Anyway, all the branching ratios of these decays are larger than $10^{-5}$, in which the maximum value can reach up to $10^{-3}$, indicating promising prospects for experimental observation.
Our predictions are consistent with other theoretical results. Furthermore,
the decays $B \to T \ell^\prime \nu_{\ell^\prime}$ are dominated by the longitudinal polarization, where the polarization fractions are close to $70\%$. Those of the decays $B \to T \tau \nu_{\tau}$ are a little smaller, where the least values coming from the decays $B_{(s)}\to D_{(s)2}^*\tau\nu_\tau$ are about $56\%$. For the decays $B \to T \ell^\prime \nu_{\ell^\prime}$, their forward-backward asymmetries $\mathcal{A}_{FB}$ are negative, while those of the decays $B \to T \tau \nu_{\tau}$ are positive. The $\mathcal{A}_{FB}$ values for the decays $B \to (a_2,f_2,K^*_2) \tau \nu_{\tau}$ are much smaller than those of the decays $B_{(s)}\to D_{(s)2}^*\tau\nu_\tau$. There exists obvious disparity about the $\mathcal{A}_{FB}$ for the decays $B \to T \tau \nu_{\tau}$ among different theoretical predictions, which remains to be clarified with further investigation.

%%%%%%%%%%%%%%%%%%%%%%%%%%%%%%%%%%%%%%%%%%%%%%%%%%%%%%%%%%%%%%%%%%%%%%%%%%%%%%%
\section*{Acknowledgment}
This work is partly supported by the National Natural Science Foundation of China under
Grant No. 11347030 and the Natural Science Foundation of Henan Province under grant
no. 232300420116, 252300421302.
%%%%%%%%%%%%%%%%%%%%%%%%%%%%%%%%%%%%%%%%%%%%%%%%%%%%%%%%%%%%%%%%%%%%%%%%%%%%%%%
\appendix

\section{Expressions of $B\to T$ transition form factors}
\label{FL2}
\be
h(q^{2}) &=& \frac{N_c}{16\pi^3}\int d x_2 d^2 p_\perp ^\prime \frac{2 h_P^\prime h_T^{\prime\prime}}{x_2 \hat{N}_1^\prime \hat{N}_1^{\prime\prime}} \bigg\{\left. x_2 m^ \prime_1+x_1 m_2+(m^\prime_1-m^{\prime\prime}_1)\frac{p_\perp^\prime \cdot q_\perp}{q^2}
+\frac{2}{w^{\prime\prime}_T}\bigg[p^{\prime2}_\perp
\right.\notag\\&&\left.
+\frac{(p_\perp^\prime \cdot q_\perp)^2}{q^2} \bigg] \right.\bigg\}
+\frac{N_{c}}{16\pi^{3}}\int d x_{2}d^{2}p^{\prime}_{\perp} \frac{2 h^{\prime}_{P}h^{\prime\prime}_{T}}{x_{2} \hat N^{\prime}_{1} \hat N^{\prime\prime}_{1}}\Big[\left.(m^{\prime}_{1}-m^{\prime\prime}_{1})(A^{(2)}_{3}+A^{(2)}_{4})\right.\notag\\
&&\left. +(m^{\prime}_{1}+ m^{\prime\prime}_{1}-2 m_{2})(A^{(2)}_{2}+A^{(2)}_{3})-m^{\prime}_{1}(A^{(1)}_{1}+A^{(1)}_{2})
\right.\notag\\&&\left.
+\frac{2}{w^{\prime\prime}_{T}}(2 A^{(3)}_{1}+2 A^{(3)}_{2}-A^{(2)}_{1}) \right.\Big],
\en

\be
b_{+}(q^{2}) &=& -\frac{N_c}{16\pi^3}\int d x_2 d^2 p_\perp ^\prime \frac{2 h_P^\prime h_T^{\prime\prime}}{x_2 \hat{N}_1^\prime \hat{N}_1^{\prime\prime}} \bigg\{\left.(x_1-x_2)(x_2 m^\prime_1+x_1 m_2)-[2 x_1 m_2+m^{\prime\prime}_1
\right.\notag\\&&\left.
+(x_2-x_1)m^\prime_1]\frac{p_\perp^\prime \cdot q_\perp}{q^2}
-2\frac{x_2 q^2+p^\prime_\perp \cdot q_\perp}{x_2 q^2 w^{\prime\prime}_T} \Big[p^\prime_\perp \cdot p^{\prime\prime}_\perp+(x_1 m_2+x_2 m^\prime_1)
\right.\notag\\&&\left.
(x_1 m_2-x_2 m^{\prime\prime}_1) \Big] \right.\bigg\}
+\frac{N_{c}}{16\pi^{3}}\int d x_{2}d^{2}p^{\prime}_{\perp} \frac{h^{\prime}_{P}h^{\prime\prime}_{T}}{x_{2} \hat N^{\prime}_{1} \hat N^{\prime\prime}_{1}}\bigg\{\left. 8(m_{2}-m^{\prime}_{1})
\right.\notag\\&&\left.
(A^{(3)}_{3}+2 A^{(3)}_{4}+A^{(3)}_{5}) -2 m^{\prime}_{1}(A^{(1)}_{1}+A^{(1)}_{2})+4(2m^{\prime}_{1}- m^{\prime\prime}_{1}-2 m_{2})(A^{(2)}_{2}+A^{(2)}_{3})
\right.\notag\\&&\left.
+2(m^{\prime}_{1}+ m^{\prime\prime}_{1})(A^{(2)}_{2}+2 A^{(2)}_{3}+A^{(2)}_{4})
+\frac{2}{w^{\prime\prime}_{T}}\Big[2[M^{\prime 2}+M^{\prime\prime 2}-q^{2}
\right.\notag\\&&\left.
+2(m^{\prime}_{1}-m_{2})(m^{\prime\prime}_{1}+m_{2})](A^{(3)}_{3}+2 A^{(3)}_{4}+A^{(3)}_{5}-A^{(2)}_{2}-A^{(2)}_{3})
 \right.\notag\\&&\left.
 +[q^{2}-\hat N^{\prime}_{1}- \hat N^{\prime\prime}_{1}-(m^{\prime}_{1}+ m^{\prime\prime}_{1})^{2}](A^{(2)}_{2}+2 A^{(2)}_{3}+A^{(2)}_{4}-A^{(1)}_{1}-A^{(1)}_{2})\Big] \right. \bigg\},
 \en

 \be
 b_{-}(q^{2}) &=& -\frac{N_c}{16\pi^3}\int d x_2 d^2 p_\perp ^\prime \frac{h_P^\prime h_T^{\prime\prime}}{x_2 \hat{N}_1^\prime \hat{N}_1^{\prime\prime}} \bigg\{\left.2(2 x_1-3)(x_2 m^\prime_1+x_1 m_2)-8(m^\prime_1-m_2)
 \right.\notag\\&&\left.
 \bigg[p^{\prime2}_\perp+2\frac{(p_\perp^\prime \cdot q_\perp)^2}{q^4} \bigg]
-[(14-12 x_1)m^\prime_1-2m^{\prime\prime}_1-(8-12 x_1)m_2]\frac{p_\perp^\prime \cdot q_\perp}{q^2}
\right.\notag\\&&\left.
+\frac{4}{w^{\prime\prime}_T}\bigg([M^{\prime2}+M^{\prime\prime2}-q^2+2(m^\prime_1-m_2)(m^{\prime\prime}_1+m_2)](A^{(2)}_3+A^{(2)}_4-A^{(1)}_2)
\right.\notag\\&&\left.
+Z_2(3 A^{(1)}_2-2 A^{(2)}_4-1)+\frac{1}{2}[x_1(q^2+q \cdot P)-2 M^{\prime2}-2 p^\prime_\perp \cdot q_\perp-2 m^\prime_1(m^{\prime\prime}_1+m_2)
\right.\notag\\&&\left.
-2 m_2(m^\prime_1-m_2)](A^{(1)}_1+A^{(1)}_2-1)+q \cdot P \bigg[p^{\prime2}_\perp+2\frac{(p_\perp^\prime \cdot q_\perp)^2}{q^4} \bigg](4 A^{(1)}_2-3) \right.\bigg)\bigg\}\left.
\right.\notag\\&&\left.
+\frac{N_{c}}{16\pi^{3}}\int d x_{2}d^{2}p^{\prime}_{\perp} \frac{h^{\prime}_{P}h^{\prime\prime}_{T}}{x_{2} \hat N^{\prime}_{1} \hat N^{\prime\prime}_{1}}
 \bigg\{8(m_2-m^{\prime}_{1})(A^{(3)}_{4}+2 A^{(3)}_{5}+A^{(3)}_{6}) -6 m^{\prime}_{1}(A^{(1)}_{1}+A^{(1)}_{2})
 \right.\notag\\&&\left.
 +4(2 m^{\prime}_{1}- m^{\prime\prime}_{1}-m_{2})(A^{(2)}_{3}+A^{(2)}_{4}) +2(3 m^{\prime}_{1}+ m^{\prime\prime}_{1}-2 m_{2})(A^{(2)}_{2}+2 A^{(2)}_{3}+A^{(2)}_{4})
 \right.\notag\\&&\left.
 +\frac{2}{w^{\prime\prime}_{T}}\Big[2[M^{\prime 2}+M^{\prime\prime 2}-q^{2}+2(m^{\prime}_{1}-m_{2})(m^{\prime\prime}_{1}+m_{2})](A^{(3)}_{4}+2 A^{(3)}_{5}+A^{(3)}_{6}-A^{(2)}_{3}-A^{(2)}_{4}) \right.\notag\\&&\left. +2 Z_{2}(3 A^{(2)}_{4}-2 A^{(3)}_{6}-A^{(1)}_{2})+2
 \frac{q\cdot P}{q^2}(6 A^{(1)}_{2}A^{(2)}_{1}-6 A^{(1)}_{2}A^{(3)}_{2}+\frac{2}{q^2}(A^{(2)}_{1})^{2}-A^{(2)}_{1}) \right.\notag\\&&\left. +[q^{2}-2 M^{\prime 2}+\hat N^{\prime}_{1}-\hat N^{\prime\prime}_{1}-(m^{\prime}_{1}+m^{\prime\prime}_{1})^{2}+2(m^{\prime}_{1}-m_{2})^2] \right.\notag\\&&\left.
 \times(A^{(2)}_{2}+2 A^{(2)}_{3}+A^{(2)}_{4}-A^{(1)}_{1}-A^{(1)}_{2})\Big]  \right.\bigg\},\\
k(q^{2}) &=& m_{B_{(s)}}m_{T}(1+\frac{m^{2}_{B_{(s)}}+m^{2}_{T}-q^{2}}{2m_{B_{(s)}}m_{T}})\Big[h(q^{2})-\frac{1}{2}b_{+}(q^{2})
+\frac{1}{2}b_{-}(q^{2})\Big],
\en
where the definations of $w^{\prime\prime}_T$ and the vertex functions $h^\prime_P$ and $h^{\prime\prime}_T$ have been given in the previous section. The explicit expressions of $Z_2$, $A^{(j)}_i$ can be found in Ref. \cite{zhangzq, cheng h y 2004}.

\section{The $B\to T$ transition form factors}
\label{FL}
\begin{table}[H]
\caption{The $B \to T$ transition form factors obtained within the CLFQM. The uncertainties mainly arise from the decay constants of the initial and final state mesons.}
\begin{center}
\scalebox{0.6}{
\begin{tabular}{|c|c|cccc|c|c|cccc|}
\hline\hline
Transition&Reference&$\quad V\quad$&$\quad A_{0}\quad$&$\quad A_{1}\quad$&$\quad A_{2}\quad$&Transition&Reference&$\quad V\quad$&$\quad A_{0}\quad$&$\quad A_{1}\quad$&$\quad A_{2}\quad$\\
\hline
$B\to\ a_{2}$&This work&$0.25^{+0.02+0.00}_{-0.01-0.01}$&$0.21^{+0.03+0.01}_{-0.03-0.01}$&$0.19^{+0.02+0.00}_{-0.01-0.01}$&$0.17^{+0.01+0.00}_{-0.00-0.01}$&$B\to\ K^{*}_{2}$&This work&$0.27^{+0.01+0.02}_{-0.02-0.01}$&$0.22^{+0.03+0.02}_{-0.03-0.02}$&$0.21^{+0.02+0.02}_{-0.02-0.01}$&$0.19^{+0.01+0.00}_{-0.01-0.01}$\\
\hline
$$&LFQM \cite{chang-qin:2112}&$0.24$&$0.21$&$0.19$&$0.17$&$$&LFQM \cite{chang-qin:2112}&$0.28$&$0.24$&$0.22$&$0.20$\\
$$&CLFQM \cite{cheng-haiyang:1010}&$0.28$&$0.24$&$0.21$&$0.19$&$$&CLFQM \cite{cheng-haiyang:1010}&$0.28$&$0.25$&$0.21$&$0.19$\\
$$&PQCD \cite{wang-wei:1008}&$0.18$&$0.18$&$0.11$&$0.06$&$$&PQCD \cite{wang-wei:1008}&$0.21$&$0.18$&$0.13$&$0.08$\\
$$&LCSR \cite{zuo 21}&$0.27$&$0.27$&$0.19$&$0.14$&$$&LCSR \cite{zuo 21}&$0.25$&$0.29$&$0.18$&$0.10$\\
$$&LCSR \cite{kcyang 2011}&$0.18$&$0.21$&$0.14$&$0.09$&$$&LCSR \cite{kcyang 2011}&$0.16$&$0.25$&$0.14$&$0.05$\\
$$&LCSR \cite{Aliev 19}&$0.18$&$0.30$&$0.16$&$0.07$&$$&LCSR \cite{Aliev 19}&$0.22$&$0.30$&$0.19$&$0.11$\\
$$&QCDSR \cite{Khosravi 16}&$0.13$&$0.26$&$0.11$&$0.09$&$$&ISGW2 \cite{scora 95,sharma 10}&$-$&$-0.17$&$-0.38$&$-0.53$\\
$$&ISGW2 \cite{scora 95,sharma 10}&$0.32$&$0.20$&$0.16$&$0.14$&$$&LFQM \cite{cheng h y 2004}&$0.28$&$0.26$&$-0.01$&$-0.21$\\
$$&LFQM \cite{cheng h y 2004}&$0.28$&$0.20$&$-0.03$&$-0.17$&$$&$$&$$&$$&$$&$$\\
\hline
$B\to\ f_{2}$&This work&$0.24^{+0.02+0.01}_{-0.02-0.01}$&$0.22^{+0.03+0.01}_{-0.03-0.02}$&$0.19^{+0.02+0.01}_{-0.02-0.00}$&$0.17^{+0.01+0.01}_{-0.01-0.00}$&$B\to\ D^{*}_{2}$&This work&$0.72^{+0.05+0.02}_{-0.05-0.02}$&$0.69^{+0.06+0.03}_{-0.06-0.02}$&$0.64^{+0.05+0.01}_{-0.05-0.01}$&$0.54^{+0.03+0.00}_{-0.02-0.01}$\\
\hline
$$&PQCD \cite{wang-wei:1008}&$0.12$&$0.13$&$0.08$&$0.04$&$$&LFQM \cite{chang-qin:2112}&$0.75$&$0.64$&$0.63$&$0.58$\\
$$&LCSR \cite{R.Khosravi:1503}&$0.30$&$0.22$&$0.17$&$0.11$&$$&$$&$$&$$&$$&$$\\
$$&LCSR \cite{zuo 21}&$0.18$&$0.18$&$0.14$&$0.11$&$$&$$&$$&$$&$$&$$\\
$$&LCSR \cite{kcyang 2011}&$0.18$&$0.20$&$0.14$&$0.10$&$$&$$&$$&$$&$$&$$\\
$$&LCSR \cite{Aliev 19}&$0.11$&$0.20$&$0.10$&$0.04$&$$&$$&$$&$$&$$&$$\\
$$&QCDSR \cite{Khosravi 16}&$0.12$&$0.24$&$0.10$&$0.09$&$$&$$&$$&$$&$$&$$\\
$$&ISGW2 \cite{scora 95,sharma 10}&$0.32$&$0.20$&$0.16$&$0.14$&$$&$$&$$&$$&$$&$$\\
\hline\hline
$B_{s}\to\ K_{2}^{*}$&This work&$0.23^{+0.02+0.01}_{-0.00-0.02}$&$0.22^{+0.03+0.03}_{-0.02-0.02}$&$0.18^{+0.02+0.01}_{-0.01-0.01}$&$0.16^{+0.00+0.01}_{-0.01-0.00}$&$B_s\to\ f_{2}^\prime$&This work&$0.31^{+0.02+0.01}_{-0.01-0.01}$&$0.31^{+0.03+0.02}_{-0.01-0.01}$&$0.25^{+0.02+0.00}_{-0.00-0.04}$&$0.21^{+0.01+0.01}_{-0.01-0.00}$\\
\hline
$$&LFQM  \cite{chang-qin:2112}&$0.23$&$0.18$&$0.17$&$0.16$&$$&LFQM \cite{chang-qin:2112}&$0.32$&$0.26$&$0.24$&$0.22$\\
$$&PQCD \cite{wang-wei:1008}&$0.18$&$0.15$&$0.11$&$0.07$&$$&PQCD \cite{wang-wei:1008}&$0.20$&$0.16$&$0.12$&$0.09$\\
$$&LCSR \cite{zuo 21}&$0.25$&$0.29$&$0.18$&$0.10$&$$&LCSR \cite{zuo 21}&$0.25$&$0.32$&$0.18$&$0.08$\\
$$&LCSR \cite{kcyang 2011}&$0.15$&$0.22$&$0.12$&$0.05$&$$&LCSR \cite{kcyang 2011}&$0.15$&$0.25$&$0.13$&$0.03$\\
$$&QCDSR \cite{Khosravi 16}&$0.13$&$0.23$&$0.10$&$0.05$&$$&ISGW2 \cite{scora 95,sharma 10}&$-$&$-0.26$&$-0.45$&$-0.59$\\
$$&ISGW2 \cite{scora 95,sharma 10}&$-$&$-0.27$&$-0.39$&$-0.47$&$$&$$&$$&$$&$$&$$\\
\hline
$B_s\to\ D_{s 2}^{*}$&This work&$0.84^{+0.05+0.02}_{-0.02-0.02}$&$0.84^{+0.06+0.02}_{-0.01-0.03}$&$0.74^{+0.04+0.01}_{-0.01-0.02}$&$0.57^{+0.02+0.00}_{-0.01-0.00}$&$$&$$&$$&$$&$$&$$\\
\hline
$$&LFQM \cite{chang-qin:2112}&$0.95$&$0.76$&$0.74$&$0.66$&$$&$$&$$&$$&$$&$$\\
$$&CLFQM \cite{zhao-zhenxing:2504}&$0.81$&$0.69$&$0.68$&$0.67$&$$&$$&$$&$$&$$&$$\\
\hline\hline
\end{tabular}\label{BtoT}
}
\end{center}
\end{table}

\begin{figure}[H]
\vspace{0.32cm}
 \centering
  \subfigure[]{\includegraphics[width=0.30\textwidth]{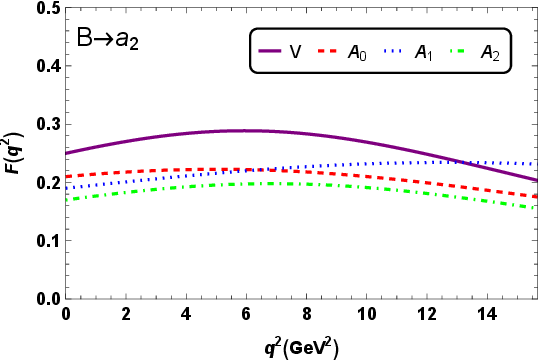}\quad}
  \subfigure[]{\includegraphics[width=0.30\textwidth]{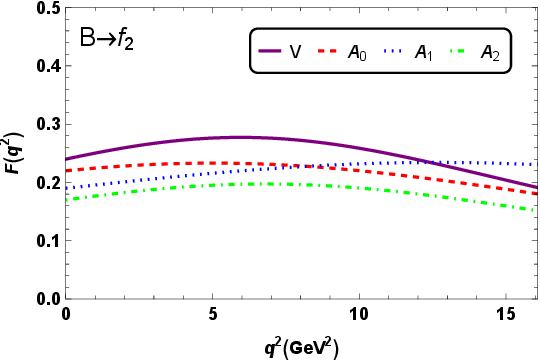}\quad}
  \subfigure[]{\includegraphics[width=0.30\textwidth]{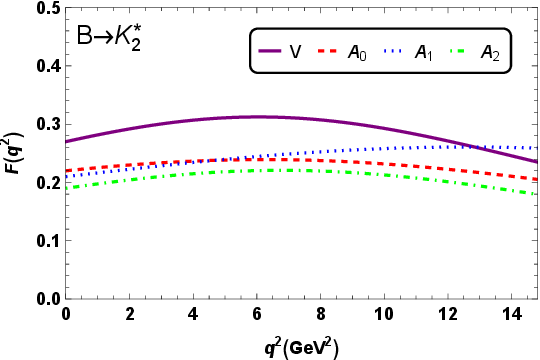}}\\
  \subfigure[]{\includegraphics[width=0.30\textwidth]{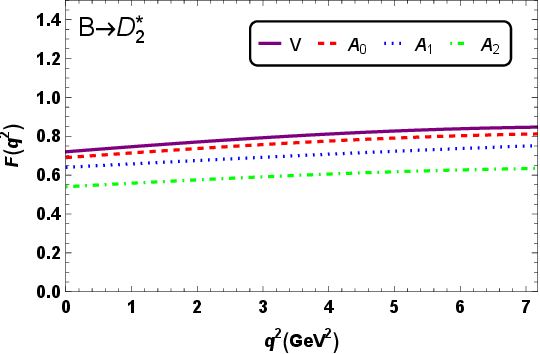}\quad}
  \subfigure[]{\includegraphics[width=0.30\textwidth]{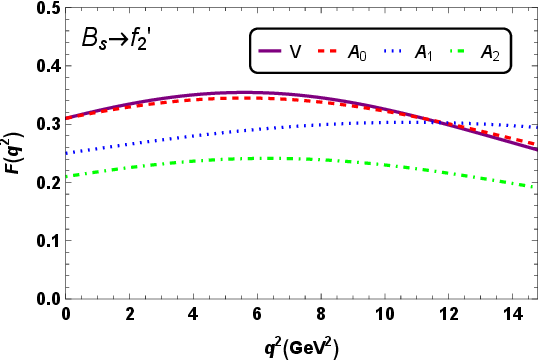}\quad}
  \subfigure[]{\includegraphics[width=0.30\textwidth]{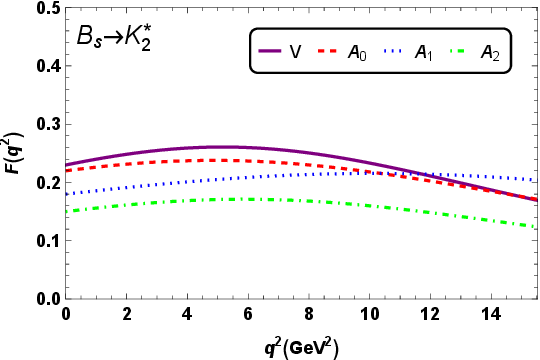}}\\
  \subfigure[]{\includegraphics[width=0.30\textwidth]{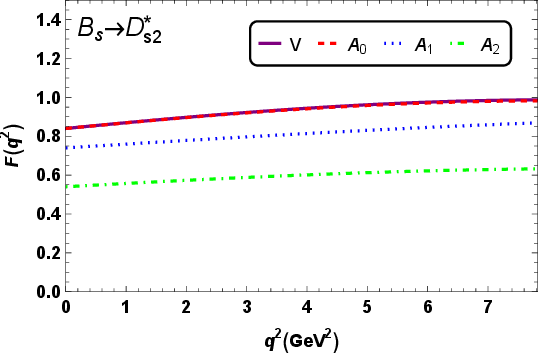}}
  \caption{Form factors $V(q^2)$, $A_0(q^2)$, $A_1(q^2)$ and $A_2(q^2)$ of the transitions $B\to a_2$ (a), $B\to f_2$ (b), $B\to K^*_2$ (c), $B\to D^*_2$ (d), $B_s\to f^\prime_2$ (e), $B_s\to K^*_2$ (f) and $B_s\to D^*_{s2}$ (g).}
  \label{BT1}
\end{figure}

%%%%%%%%%%%%%%%%%%%%%%%%%%%%%%%%%%%%%%%%%%%%%%%%%%%%%%%%%%%%%%%%%%%%%%%%
%                               references
%%%%%%%%%%%%%%%%%%%%%%%%%%%%%%%%%%%%%%%%%%%%%%%%%%%%%%%%%%%%%%%%%%%%%%%%

\end{document}